\documentclass[onecolumn]{aastex63}
\usepackage{graphicx}
\usepackage{hyperref}
\usepackage{amsmath}

\received{XXX}
\revised{YYY}
\accepted{ZZZ}

\submitjournal{ApJS}

\newcommand{\efRoundThree}{}

\newcommand{\psrchive}{{\tt psrchive}}
\newcommand{\pp}{{\tt PulsePortraiture}}

\begin{document}
\title{Modeling the Morphology of Fast Radio Bursts and Radio Pulsars with {\tt fitburst}}

\shorttitle{Modeling Radio Pulses with {\tt fitburst}}

\author[0000-0001-8384-5049]{E. Fonseca}
    \affiliation{Department of Physics and Astronomy, West Virginia University, Morgantown, WV 26506-6315, USA}
    \affiliation{Center for Gravitational Waves and Cosmology, Chestnut Ridge Research Building, Morgantown, WV 26505, USA}

\author[0000-0002-4795-697X]{Z. Pleunis}
    \affiliation{Dunlap Institute for Astronomy \& Astrophysics, University of Toronto, 50 St.~George Street, Toronto, ON M5S 3H4, Canada}

\author[0000-0002-2349-3341]{D. Breitman}
    \affiliation{Scuola Normale Superiore, Piazza dei Cavalieri 7, I-56126 Pisa, Italy}

\author[0000-0003-3154-3676]{K.~R. Sand}
    \affiliation{Department of Physics, McGill University, 3600 rue University, Montr\'eal, QC H3A 2T8, Canada}
    \affiliation{Trottier Space Institute, McGill University, 3550 rue University, Montr\'eal, QC H3A 2A7, Canada}

\author[0009-0008-6166-1095]{B. Kharel}
    \affiliation{Department of Physics and Astronomy, West Virginia University, Morgantown, WV 26506-6315, USA}
    \affiliation{Center for Gravitational Waves and Cosmology, Chestnut Ridge Research Building, Morgantown, WV 26505, USA}

\author[0000-0001-8537-9299]{P.~J. Boyle}
    \affiliation{Department of Physics, McGill University, 3600 rue University, Montr\'eal, QC H3A 2T8, Canada}
    \affiliation{Trottier Space Institute, McGill University, 3550 rue University, Montr\'eal, QC H3A 2A7, Canada}

\author[0000-0002-1800-8233]{C. Brar}
    \affiliation{Department of Physics, McGill University, 3600 rue University, Montr\'eal, QC H3A 2T8, Canada}
    \affiliation{Trottier Space Institute, McGill University, 3550 rue University, Montr\'eal, QC H3A 2A7, Canada}

\author[0000-0001-5553-9167]{U. Giri}
    \affiliation{Department of Physics, University of Wisconsin-Madison, 1150 University Ave, Madison, WI 53706, USA}

\author[0000-0001-9345-0307]{V.~M. Kaspi}
    \affiliation{Department of Physics, McGill University, 3600 rue University, Montr\'eal, QC H3A 2T8, Canada}
    \affiliation{Trottier Space Institute, McGill University, 3550 rue University, Montr\'eal, QC H3A 2A7, Canada}

\author[0000-0002-4279-6946]{K.~W. Masui}
    \affiliation{MIT Kavli Institute for Astrophysics and Space Research, Massachusetts Institute of Technology, 77 Massachusetts Ave, Cambridge, MA 02139, USA}
    \affiliation{Department of Physics, Massachusetts Institute of Technology, 77 Massachusetts Ave, Cambridge, MA 02139, USA}

\author[0000-0001-8845-1225]{B.~W. Meyers}
    \affiliation{International Centre for Radio Astronomy Research, Curtin University, Bentley, WA 6102, Australia}

\author[0000-0003-3367-1073]{C. Patel}
    \affiliation{Department of Physics, McGill University, 3600 rue University, Montr\'eal, QC H3A 2T8, Canada}
    \affiliation{Trottier Space Institute, McGill University, 3550 rue University, Montr\'eal, QC H3A 2A7, Canada}

\author[0000-0002-7374-7119]{P. Scholz}
    \affiliation{Dunlap Institute for Astronomy \& Astrophysics, University of Toronto, 50 St.~George Street, Toronto, ON M5S 3H4, Canada}
    \affiliation{Department of Physics and Astronomy, York University, 4700 Keele Street, Toronto, ON MJ3 1P3, Canada}

\author[0000-0002-2088-3125]{K. Smith}
    \affiliation{Perimeter Institute for Theoretical Physics, 31 Caroline Street N, Waterloo, ON N25 2YL, Canada}

\begin{abstract}
We present a framework for modeling astrophysical pulses from radio pulsars and fast radio bursts (FRBs). This framework, called {\tt fitburst}, generates synthetic representations of dynamic spectra that are functions of several physical and heuristic parameters; the heuristic parameters can nonetheless accommodate a vast range of distributions in spectral energy. {\tt fitburst} is designed to optimize the modeling of features induced by effects that are intrinsic and extrinsic to the emission mechanism, including the magnitude and frequency dependence of pulse dispersion and scatter-broadening. {\tt fitburst} removes intra-channel smearing through two-dimensional upsampling, and can account for phase wrapping of ``folded" signals that are typically acquired during pulsar-timing observations. We demonstrate the effectiveness of {\tt fitburst} in modeling data containing pulsars and FRBs observed with the Canadian Hydrogen Intensity Mapping Experiment (CHIME) telescope.
\end{abstract}


\section{Introduction}
\label{sec:intro}

One of the key observables in radio-transient astrophysics is the underlying shape of the detected spectral energy over time and electromagnetic frequency, which we hereafter refer to as ``morphology." In the context of radio pulsars and fast radio bursts (FRBs), aspects of the observed pulse morphology often serve as tools for probing various physical phenomena that are both intrinsic or extrinsic to the underlying emission mechanism. For example, the morphology of radio pulsars usually exhibits a high degree of constancy that forms a central tenet to pulsar timing; this stability allows for accurate and precise measurements of arrival times, and for the resolution of time-domain effects with broad applications \citep[e.g.,][]{lk12}. By contrast, the emerging field of FRBs is providing strong evidence for delineations between distinct types of FRB sources, based purely on morphology, that may reflect underlying differences in progenitor populations and/or emission physics \citep[e.g.,][]{pgk+21}.

The primary quantity that characterizes single-pulse morphology is the {\it dynamic spectrum}, $D(\nu,t)$, which maps source intensity across different values of time ($t$) and electromagnetic frequency ($\nu$). For radio pulsars, the periodic nature of the emission leads to time-averaged morphology being defined across $\nu$ and rotational phase $\phi \equiv \phi(t')$ where $t'$ is the timescale of rotation, meaning that the ``phase-resolved dynamic spectrum" $D \equiv D(\nu, \phi$) for pulsars.\footnote{This distinction is important because pulsar literature often uses the phrase ``dynamic spectrum" to mean $D = \int D'(\nu,t, \phi)d\phi$, i.e., a ``true" dynamic spectrum that is marginalized over morphology. For the purposes of this present work, we define $D = \int D'(\nu,t,\phi)dt$ in order to preserve morphology for analysis.} In either case, $D$ encodes information that is intrinsic and extrinsic to the radiation source, arising from the generation and propagation of radio emission. 

Prior studies have characterized the effects that impact observed morphology using frameworks which model a subset of physical effects while marginalizing over other data dimensions and/or avoiding treatment of the underlying spectral energy distribution (SED), both of which can impact parameter estimation \citep[e.g.,][]{rav19,aal+21,jan22}. Moreover, the rise of broadband radio telescope receivers is producing wide-band data that allow for simultaneous evaluation of SEDs and parameters that govern propagation effects. In this work, we describe a framework that generates broadband models of $D$ in terms of physical and phenomenological parameters that encapsulate all aspects of morphology and for bursts of different SED shapes. Our framework, hereafter referred to as {\tt fitburst}\footnote{\url{https://chimefrb.github.io/fitburst/}}, is a substantial expansion of the initial spectrum model used to study FRB 20110523A \citep{mls+15}.

The goal of {\tt fitburst} is to generate synthetic, noiseless representations of $D$ that are functions of several macroscopic quantities which can then be studied through subsequent statistical analyses. We consider this work complementary to recent efforts in constructing broadband ``portrait" models of $D$ for pulsar-timing data \citep[e.g.,][]{pdr14,ldc+14,lkd+17,pen19}, as our methods seek to characterize morphological structures while others assume prior knowledge of the underlying pulse shape. The {\tt fitburst} algorithm has been under development over the past few years and initial descriptions of the algorithm and results obtained with it have been described across previous works \citep[e.g.,][]{mls+15,chimefrb20,pgk+21,afm+23,sbm+23}.

In this work, we describe {\tt fitburst} in its fullest generality, publicly release the codebase and, for the first time, present a detailed description of its exact estimation of the covariance matrix. In Section \ref{sec:models}, we describe the theoretical basis of our models for $D$. Afterwards, we outline the optimization procedure for smoothly varying and stochastic SEDs in Section \ref{sec:method}. In Section \ref{sec:analysis}, we demonstrate the effectiveness of {\tt fitburst} when applied to different types of data acquired with the Canadian Hydrogen Intensity Mapping Experiment \citep[CHIME;][]{chime22} telescope. We conclude our work with discussions regarding various technical aspects of {\tt fitburst} in Section \ref{sec:discussion}.

\section{Models of Dynamic Spectra}
\label{sec:models}

In order to construct a morphological model, we assumed that any pulse within a dynamic spectrum can be intrinsically characterized by its temporal width ($\sigma$) and mean time of arrival at the detector ($t_0$); the underlying temporal shape can then be chosen to reflect Gaussian or other forms, and model comparisons can be enacted to discern which of these shapes best describe the data. In what follows, we assumed that any distinct pulse present in $D$ is intrinsically Gaussian in its temporal shape. The frequency dependence of a pulse observed in $D(\nu, t)$, however, is subject to greater modeling uncertainty due to observed differences in SEDs between pulsars and FRBs. 

At an extrinsic level, pulse dispersion and broadening from multi-path scattering will induce frequency-dependent variations in observed arrival times and morphology of all visible pulse structure \citep[e.g.,][]{cor02}. These respective plasma effects are often characterized by a ``dispersion measure" (DM) and a timescale of one-sided, exponential scattering-broadening of an intrinsically Gaussian profile ($\tau$), though deviations from ideal scattering geometries can lead to complex pulse morphology \citep[e.g.,][]{gk16}. The ``thin-screen" formalism of scatter-broadening, widely used in pulsar and FRB astronomy, predicts that $\tau \propto \nu^{-4}$, while propagation through cold plasma leads to a time delay $\Delta t \propto \nu^{-2}$. 

Using these assumptions, we defined the model of a single pulse to consist of three distinct parts: a frequency-dependent SED ($F_k$); a dispersed temporal shape of the pulse ($T_{kn}$); and a global amplitude ($A$). The $k$ and $n$ indices denote labels for discrete frequency channels ($\nu_k$) and time bins ($t_n$), respectively. For a spectrum with two or more distinct pulses, the full model ($M_{kn}$) is a superposition of models for each component labeled with an index $l$. The analytic expression for $T_{kn,l}$ depends on whether scatter-broadening is measurable \citep[e.g.,][]{mck14}; if scatter-broadening is significant, then the full model of a dynamic spectrum with $N$ scatter-broadened, intrinsically Gaussian pulses is written as $M_{kn}=\sum_l^NA_lF_{k,l}T_{kn,l}$, where 

\begin{align}
    \label{eq:amp_def}
    A_l &= 10^{\alpha_l}, \\
    \label{eq:sed_def}
    F_{k,l} &= \bigg(\frac{\nu_k}{\nu_r}\bigg)^{\gamma_l+\beta_l\ln(\nu_k/\nu_r)}, \\
    \label{eq:t_def}
    T_{kn,l} &= \bigg(\frac{\nu_k}{\nu_r}\bigg)^{-\delta}\exp{\bigg[\frac{\sigma_l^2}{2\tau_k^2}-\frac{(t_{kn}-t_{0,l})}{\tau_k}\bigg]}\bigg\{1+{\rm erf}\bigg[\frac{t_{kn}-(t_{0,l}+\sigma_l^2/\tau_k)}{\sigma_l\sqrt{2}}\bigg]\bigg\} \textrm{ for $\tau_r > 0$ ms}.
\end{align}

\noindent The scatter-broadening timescale $\tau_k$ and dispersed timeseries $t_{kn}$ can be expressed in terms of a (fixed) reference frequency $\nu_r$,

\begin{align}
    \label{eq:tau_def}
    \tau_k &= \tau_r\bigg(\frac{\nu_k}{\nu_r}\bigg)^\delta \\ 
    \label{eq:tkn_def}
    t_{kn} &= t_n - k_{\rm DM}{\rm DM}\big(\nu_k^\epsilon - \nu_r^\epsilon\big),
\end{align}

\noindent where $\tau_r$ is the scattering timescale at $\nu_r$ while \{$\delta$, $\epsilon$\} are the exponents of the frequency dependence for scatter-broadening and dispersion, respectively. The ``dispersion constant" $k_{\rm DM} = (2.41\times10^{-4})^{-1}$ pc$^{-1}$ cm$^3$ GHz$^2$ s in keeping with community convention.\footnote{$k_{\rm DM}$ is a collection of physical constants whose definitions have improved as the field of pulsar astronomy has progressed, leading to the awkward circumstance where the physically accurate value of $k_{\rm DM}$ now differs significantly from the community standard \citep[e.g.,][]{kul20}. {\tt fitburst} sets the value of $k_{\rm DM}$ as a configurable parameter. We nonetheless urge caution against changing its value as many decades of pulsar literature, as well as ongoing high-precision timing experiments, are based on and sensitive to the antiquated value.} We replaced Equation \ref{eq:t_def} with a Gaussian shape of mean $t_{0,l}$ and width $\sigma_l$ in the case where scatter-broadening cannot be resolved at the native time resolution of $D_{kn}$:

\begin{equation}
    T_{kn,l} = \exp\bigg[\frac{1}{2}\bigg(\frac{t_{kn}-t_{0,l}}{\sigma_l}\bigg)^2\bigg] \textrm{ for $\tau_r = 0$ ms}
\end{equation}

\noindent A summary of symbols defined and used throughout this work is presented in Table \ref{tab:symbols}.

\begin{deluxetable}{lr}
    \label{tab:symbols}
    \tablecolumns{2}
    \tablehead{\colhead{Parameter} & \colhead{Description}}
    \tablecaption{Variables defined and used throughout this work.}
    \tablenum{1}
    \startdata
    $\alpha_l$ & Base-10 exponent of amplitude for burst componnent $l$ \\
    $\beta_l$ & Spectral running for burst component $l$ \\
    $\gamma_l$ & Spectral index for burst component $l$ \\
    $\delta$ & Exponent of frequency dependence for scatter-broadening \\
    $\epsilon$ & Exponent of frequency dependence for pulse dispersion \\
    $\nu_k$ & Electromagnetic frequency for channel $k$ \\
    $\nu_r$ & Reference (or ``pivot") electromagnetic frequency \\
    $\sigma_l$ & Temporal width of burst component $l$ \\
    $\tau_k$ & Timescale of scattering-broadening at $\nu_k$ \\
    $\tau_r$ & Timescale of scattering-broadening at $\nu_r$ \\
    $\chi^2$ & Goodness-of-fit statistic, the weighted sum of squared residuals \\
    $\Delta \nu$ & Total receiver bandwidth \\
    $\Delta \nu_k$ & Frequency resolution of channel $k$ \\
    $\Delta t_n$ & Time resolution of sample $n$ \\
    $\Delta t_{kn}$ & ``Full" temporal width of burst at channel $k$ \\
    $t_n$ & Timestamp for sample $n$ of input timeseries \\
    $t_{kn}$ & $t_n$ corrected for dispersion at $\nu_k$ \\
    $D$ & Dynamic spectrum (``phase-resolved" for pulsars) \\
    $D_{kn}$ & Value of $D$ at $\nu_k$ and $t_n$ \\
    DM & Dispersion measure \\
    $M_{kn}$ & Model of $D_{kn}$ \\
    $N$ & Number of burst components \\
    $N_t$ & Number of time samples \\
    $N_\nu$ & Number of electromagnetic frequency channels \\
    $N_{\rm dof}$ & Number of degrees of freedom in a {\tt fitburst} fit \\
    $N_{\rm fit}$ & Number of variable model parameters in a {\tt fitburst} fit \\
    $T_{kn,l}$ & Temporal profile at $\nu_k$ and $t_n$ for burst component $l$ \\
    \enddata
\end{deluxetable}

\subsection{Running power-law SED for Radio Pulses}
\label{subsec:rpl}

As indicated by Equation \ref{eq:sed_def}, we assumed the SED to follow a ``running" power-law (RPL) model, where the spectral index $\kappa_l = \gamma_l+\beta_l\ln(\nu/\nu_r)$ depends on $\nu$.\footnote{The term ``running" is borrowed from nomenclature used in cosmology when characterizing the power spectrum of the cosmic microwave background \citep[e.g.,][]{planck20}.} Such a form can be interpreted as a first-order Taylor expansion of a spectral index $\kappa$ as a function of $\ln(\nu/\nu_r)$, where $\beta_l = [d\kappa_l/d(\ln[\nu/\nu_r])]|_{0}$. In practice, this SED model is desirable because it can adequately describe both broadband and Gaussian-like SEDs as described in later sections. 

The distinction between broadband and Gaussian-like SEDs lies in the values of \{$\beta$, $\gamma$\}: broadband signals correspond to $\beta \approx 0$ and $|\gamma| \le 10$, while Gaussian-like SEDs corresponds to large magnitudes for both parameters. The RPL model is thus ideally flexible for analyzing a diverse range of morphology with a single framework. A discussion on the relationship between \{$\beta$, $\gamma$\} and the traditional Gaussian mean and variance (for band-limited signals) is provided in Appendix \ref{sec:app_gaussian}.

\subsection{Parameterization of Amplitude}
Equation \ref{eq:amp_def} shows that the ``amplitude parameter" in {\tt fitburst} ($\alpha$) is the base-10 logarithm of the ``full" amplitude ($A$). This choice is largely motivated by the interplay between the \{$\alpha$, $\beta$, $\gamma$\} parameters and the chosen value of $\nu_r$, which is discussed in Appendix \ref{sec:app_gaussian}. For SEDs with $\beta \approx 0$, $A$ will vary by factors of 1--10 for modern values of receiver bandwidth ($\Delta\nu$) and values of $\gamma$ that are typical for radio pulsars in the Milky Way Galaxy. However, for Gaussian-like SEDs, $A$ will be considerably small in order of magnitude if $\nu_r$ is sufficiently different from the frequency of peak emission. If $A$ was chosen as the variable parameter, then an initial guess for $A$ could easily differ by many orders of magnitude from its ``true" value when applied to spectra that have $|\beta| \gg 0$ and if $\nu_r$ is not carefully chosen. This large difference leads to prolonged or failed searches for the optimal value of $A$ for pulses with Gaussian-like SEDs if $\nu_r$ is arbitrarily chosen.

One solution to this problem is for the user to choose a value of $\nu_r$ that is approximately equal to the frequency of peak emission for a pulse with Gaussian-like SED in $D_{kn}$. However, such a method will be cumbersome when seeking to model a large and morphologically diverse collection of $D_{kn}$ where pulses with Gaussian-like SEDs span different parts of an observed bandwidth. Moreover, it may be useful to obtain a large sample of \{$\beta$, $\gamma$\} values that are all referenced to the same $\nu_r$ for statistical analysis of SED variations. By choosing $\alpha$ as the fit parameter, {\tt fitburst} can estimate amplitudes of observed spectra in a coordinate phase space where linear steps in $\alpha$ correspond to logarithmic steps in $A$, and thus efficiently find the optimal value for $A$ regardless of SED shape.

\subsection{Global and per-pulse Parameters}
\label{subsec:parameters}

Equations \ref{eq:amp_def}--\ref{eq:tkn_def} define the model $M_{kn}$ in terms of two parameter subsets. The \{DM, $\tau_r$, $\delta$, $\epsilon$\} parameters govern the extrinsic plasma effects of dispersion and scatter-broadening; we assumed that these ``global" parameters apply to all astrophysical features within a given dynamic spectrum and do not vary over the timescales of observed features. The five ``per-pulse" parameters \{$\alpha_l$, $\sigma_l$, $t_{0,l}$, $\beta_l$, $\gamma_l$\} are instead uniquely defined for each of the $N$ pulses present in $D_{kn}$, regardless of the prevalence of scatter-broadening. In all cases, $\nu_r$ is considered to be a fixed parameter that can be set to any arbitrary, non-zero value. The most general form of $M_{kn}$, where scatter-broadening is assumed and all variable parameters are chosen as degrees of freedom, therefore depends on $N_{\rm fit}$ = (4+5$N$) fit parameters.

In practice, it can be beneficial to generate models where one or more of the parameters are held fixed to predetermined values. For example, and as described above, plasma theory under commonly employed assumptions predicts that $\delta=-4$ and $\epsilon=-2$. Moreover, $\tau_r=0$ ms for features where no scatter-broadening across the total bandwidth is resolvable. One or more of these fixed-parameter combinations define models that are ``nested" from the most general form of $M_{kn}$. Several such realizations of $M_{kn}$ can subsequently be compared against observed data through statistical analyses of their goodness of fit, which we discuss in later sections. We nonetheless present our model in its fullest generality, and clarify fixed-parameter assumptions when they are used during analysis. 

\begin{figure}
    \centering
    \includegraphics[scale=0.035]{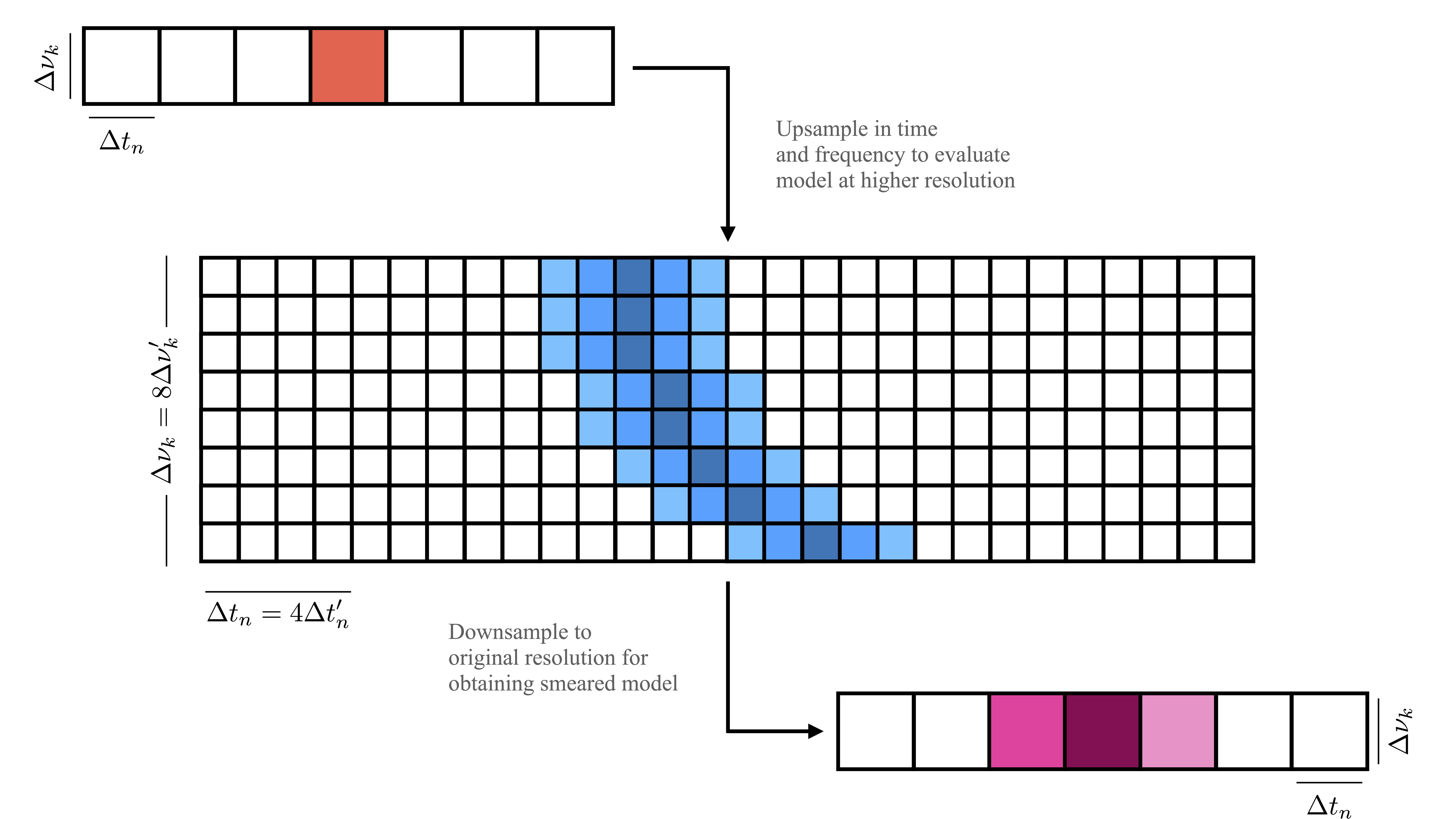}
    \caption{Calculation of dispersion smearing in {\tt fitburst} models. By default, $M_{kn}$ will inherit the original time and frequency resolution of $D_{kn}$; this resolution will yield a model for unresolved, highly-dispersed profiles as shown in red in the top-most timeseries. To account for smearing within each channel, {\tt fitburst} will first produce a high-resolution version of each channelized timeseries, upsampled along both axes by user-specified amounts. {\tt fitburst} will then evaluate the model across the upsampled grid, where intra-channel dispersion can be better resolved. (The above figure shows a dispersed model of the original timeseries in blue shading that is obtained after upsampling in frequency by a factor of 8, and in time by a factor of 4.) The high-resolution model is then downsampled to the original resolution to form a dispersion-smeared pulse, where the signal of a dispersed, intrinsically low-width pulse now appears in more than one time sample.}
    \label{fig:smearing1}
\end{figure}

\subsection{Treatment of Intra-channel Dispersion Smearing}
\label{subsec:smearing_method}

We used the $kn$-index notation above to emphasize the discretized nature of observed dynamic spectra.  A consequence of discretely sampling raw telescope data is the incoherent removal of pulse dispersion when applying Equation \ref{eq:tkn_def} to channelized data: while all channels will be de-dispersed relative to each other, each channel will suffer a smearing of the intrinsic signal evaluated over its finite bandwidth ($\Delta\nu_k$). The temporal extent of pulse smearing across a channel bandwidth $\Delta\nu_k$ can be estimated to be $\Delta t_{\rm smear} \approx \epsilon k_{\rm DM}{\rm DM}\nu^{\epsilon-1}\Delta \nu_k$. The full, ``observed" width of a pulse ($\Delta t_k$) can thus be estimated by adding the intrinsic and broadening terms in quadrature,

\begin{equation}
    \Delta t_k^2 = \Delta t_n^2 + \sigma^2 + \tau_k^2 + \Delta t_{\rm smear}^2,
\end{equation}

\noindent where $\Delta t_n$ denotes the limit in resolvable width from the native time resolution. Incoherent de-dispersion will always lead to a non-zero amount of $\Delta t_{\rm smear}$. The combination of dispersion and time/frequency resolution therefore impacts our modeling of the intrinsic pulse width $\sigma$. 

The magnitude of $\Delta t_{\rm smear}$ can nonetheless be minimized through an additional modeling procedure when generating $M_{kn}$. Our chosen method consists of two steps: first generate $M_{k'n'}$ at higher frequency and time resolutions than afforded in the observed spectrum $D_{kn}$; then downsample $M_{k'n'}$, in both dimensions and by appropriate factors, in order to find $M_{kn}$ with dimensions that match those of $D_{kn}$. A schematic of this dispersion-smearing removal is illustrated in Figure \ref{fig:smearing1}. This method of dispersion-smearing removal introduces additional computations in our modeling, but allows for model-independent estimation of both $\sigma$ and DM.

It is sometimes advantageous to manipulate raw telescope data in order to remove the deleterious effects of dispersion prior to analysis. For example, this manipulation is regularly performed for ``coherently dedispersed" timing observations of radio pulsars, where intra-channel smearing for a specified DM is completely removed prior to the formation the total-intensity dynamic spectrum \citep{hr75}. The above procedure for smearing removal can nonetheless apply for coherently dedispersed data, so long as the DM parameter instead corresponds to a value offset from the DM used for coherent dedispersion. {\tt fitburst} can accommodate these differences through proper configuration when modeling data that are coherently or incoherently dedispersed.

\section{Fitting Methods}
\label{sec:method}

We evaluate the goodness of fit between $M_{kn}$ and $D_{kn}$ using a weighted-$\chi^2$ statistic:

\begin{equation}
    \label{eq:chi2}
    \chi^2 = \sum_{k}^{N_\nu}\sum_{n}^{N_t}\bigg(\frac{D_{kn} - M_{kn}}{w_k}\bigg)^2,
\end{equation}

\noindent where $w_k$ represents a per-channel ``weight" that we define as the standard deviation in each channel, evaluated over a user-specified fraction of time. For our model described in Section \ref{sec:models}, the $\chi^2$ statistic is at most a function of the (4+5N) parameters described in Section \ref{subsec:parameters} and therefore spans a (4+5N)-dimensional phase space at maximum. The global minimum in the $\chi^2$ phase space represents the optimal fit of $M_{kn}$ to $D_{kn}$, where the Jacobian vector $\nabla\chi^2 = 0$.

Our implementation of the {\tt fitburst} model and fitting procedure is written in the Python language. We employed a weighted least-squares minimization algorithm in our implementation of {\tt fitburst}, using Equation \ref{eq:chi2} to obtain best-fit estimates of $M_{kn}$ from our observed $D_{kn}$. {\tt fitburst} can also perform ordinary least-squares fitting, where $w_k=1$, though we encourage the use of Equation \ref{eq:chi2} so that arbitrary normalization of $D_{kn}$ can be taken into account. Our implementation of the fit procedure used the {\tt least\_squares} routine in the {\tt scipy.optimize} package. In order to efficiently determine the region of best fit, we derived exact expressions for the components of $\nabla\chi^2$ and supplied it as a subroutine to {\tt least\_squares}. We also derived exact expressions for the components of the Hessian matrix, in order to estimate the covariance matrix and statistical uncertainties of all fitted parameters. For completeness, the exact form of $\nabla\chi^2$ is provided in Appendix \ref{sec:app_jacobian}.

\subsection{Amplitude-independent Fitting in Cases of Scintillation}
\label{subsec:scintillation}

The models described in Section \ref{sec:models} are most appropriate for signals with smoothly-varying SEDs across a given bandwidth. However, a common astrophysical occurrence is the stochastic variation of brightness across frequency due to interstellar scintillation \citep[e.g.,][]{ric70,cpl86}. As a consequence, the ISM imparts intensity fluctuations that cannot be adequately modeled using simple analytic functions. Direct modeling of scintillation can nonetheless be performed by estimating an amplitude for each frequency channel, though such an effort will substantially increase the number of degrees of freedom by an amount equal to the number of frequency channels ($N_\nu$), which can often exceed $N_\nu > 10^3$ for modern telescope-receiver systems.

As shown by \cite{tay92} and \cite{pdr14}, scintillation can be indirectly modeled by modifying Equation \ref{eq:chi2} to be defined for a subset of model parameters. We first note that the quantity $A_lF_{k,l}$ that defines $M_{kn}$ represents a per-channel amplitude $A_{k,l}$. In cases where scintillation is apparent, we are interested in minimizing $\chi^2$ in the subspace where all derivatives with respect to $A_{k,l}$ are already minimized. This condition corresponds to evaluating the derivative of $\chi^2$ with respect to $A_{k,l}$ and requiring that all such quantities be equal to zero; the result is a system of $N$ equations that are linear functions of the $A_{k,l}$ parameters. In the case where $N=1$, we find that 

\begin{equation}
    \label{eq:amp_scint}
    A_k = \frac{\sum_n^{N_t}D_{kn}T_{kn}}{\sum_n^{N_t}T^2_{kn}},
\end{equation}

\noindent which is consistent with the form derived by \cite{pdr14}. However, the general expression of $A_{k,l}$ for an $N$-component model is a convoluted function of $D_{kn}$ and $T_{kn,l}$ that defines a system of $N$ linear equations for each channel with index $k$. This system is most easily summarized with a recursion relation:

\begin{equation}
    \label{eq:amp_scint_general}
    \sum_m^N\sum_n^{N_t}A_{k,m}T_{kn,l}T_{kn,m} = \sum_n^{N_t}D_{kn}T_{kn,l}.
\end{equation}

The model $M_{kn}=\sum_l^NA_{k,l}T_{kn,l}$ can therefore be fully determined in an amplitude-independent manner using Equation \ref{eq:amp_scint} for $N=1$, or numerically solving Equation \ref{eq:amp_scint_general} for each value of $A_{k,l}$. While ideal for generating models in the presence of scintillation, this method introduces additional modeling computations and necessarily relinquishes the power to constrain SED parameters. This circumstance leads to the reduction of per-component fit parameters to \{$t_{0,l}$, $\sigma_l$\}, for a maximum of (4+$2N$) parameters. The DM and scatter-broadening parameters can nonetheless be determined uniquely as they define $T_{kn,l}$, regardless of the assumed SED model. 

\begin{figure}
    \centering
    \includegraphics[scale=0.58]{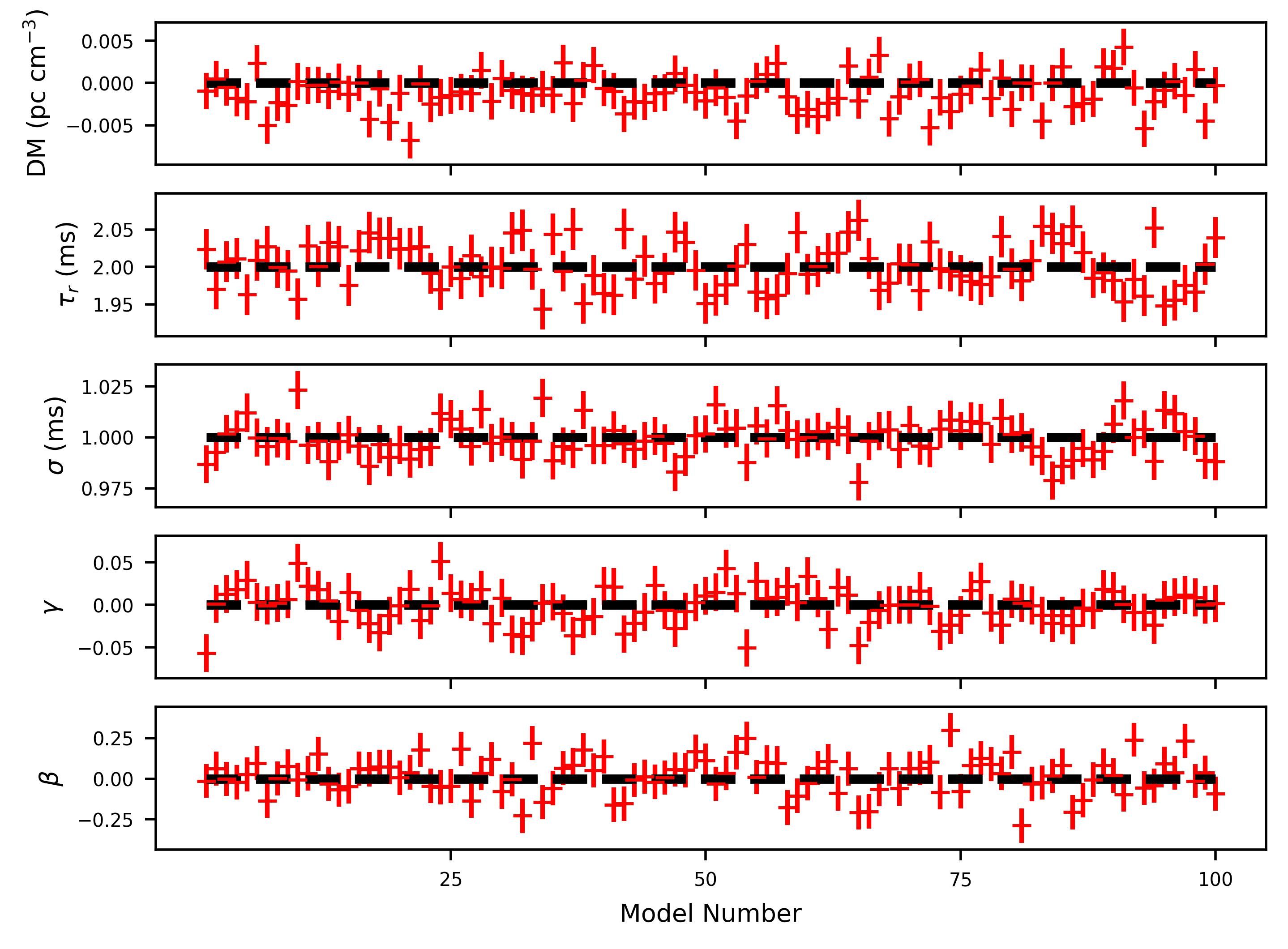}
    \includegraphics[scale=0.56]{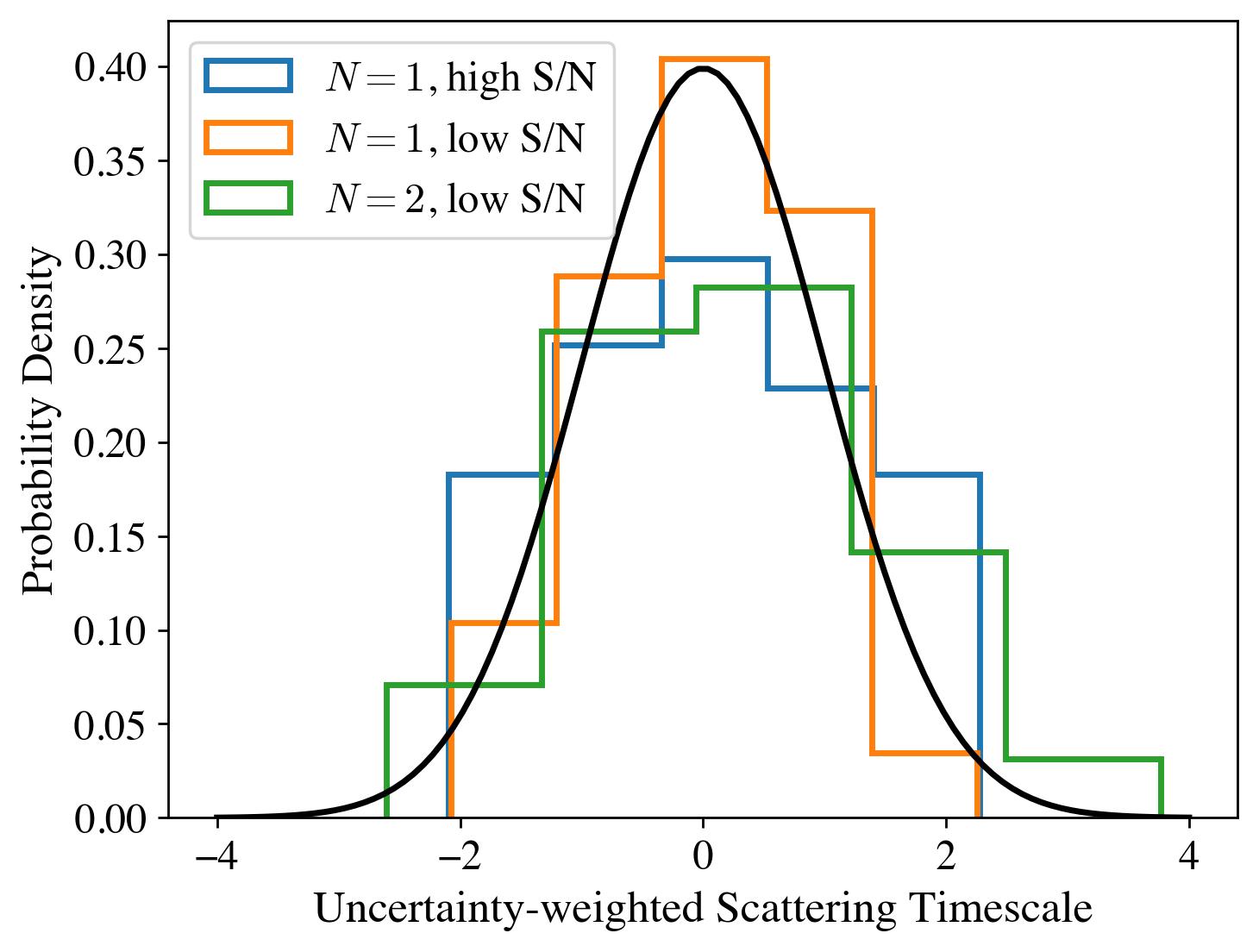}
    \caption{\efRoundThree{{(Left.)}} Best-fit measurements of {\tt fitburst} parameters derived from data simulated \efRoundThree{with} {\tt simpulse}, using an initial guess described in Section \ref{subsec:sims}. The statistical uncertainty for each point corresponds to the 68.3\% confidence interval computed by {\tt fitburst}, and black-dashed lines correspond to {\tt simpulse} input values. \efRoundThree{(Right.) Distributions of uncertainty-weighted deviations of $\tau_r$ from the mean value derived from ``high-S/N" ($\sim 200$) and ``low-S/N" ($\sim 20$) simulated bursts, as well as one-component ($N=1$) and two-component ($N=2$) bursts, each consisting of 100 simulations. All sets of simulations yield estimates of uncertainty-weighted $\tau_r$ deviates that largely reflect the normal distribution of the simulation noise (black line).}}
    \label{fig:sims}
\end{figure}

\subsection{Assessment of Viability with Simulations}
\label{subsec:sims}

In order to assure the validity of our proposed framework, we simulated a large number of dedispersed radio pulses using the independently developed {\tt simpulse} library \citep{mts+23} for analysis with {\tt fitburst}. For all simulations, we used the {\tt simpulse} library to produce coherently dedispersed dynamic spectra in the CHIME band (i.e., 400-800 MHz) with $N_\nu=1024$, $N=1$, $\sigma=1$ ms, and $\tau_r = 2$ ms at $\nu_r = 600$ MHz. Moreover, all {\tt simpulse} simulations assumed a value of $\gamma = 0$, corresponding to a flat spectrum; as {\tt simpulse} does not use an RPL model in its calculations, we expect that all {\tt fitburst} results derived from these simulations should yield $\beta=0$, i.e., no spectral running. We then created 100 realizations with a time resolution of 40 $\mu$s after adding Gaussian noise\efRoundThree{, setting the variance such that the peak S/N $\sim 200$ for each busrt}.

A set of results that demonstrate the reliability of {\tt fitburst} under different noise realizations is presented in Figure \ref{fig:sims}. For all {\tt fitburst} fits \efRoundThree{described below}, we used the weighted least-squares fitting procedure described above to \efRoundThree{optimize} all parameters \efRoundThree{against the ``high-S/N" data described above}, except \{$\delta=-4$, $\epsilon=-2$\} \efRoundThree{which were held fixed}; we applied the same initial guess \{$\alpha=0$, $\beta=0$, $\gamma=0$, $\sigma=0.5$ ms, $\tau_r=10$ ms, DM = 0 pc cm$^{-3}$\} \efRoundThree{for each execution of {\tt fitburst} and} for a fixed value of $\nu_r=600$ MHz, leading to $N_{\rm fit} = 7$ \efRoundThree{for all fits. W}e did not perform any upsampling and treated the DM as an offset parameter as all spectra were constructed to be coherently dedispersed. Each panel shows the best-fit value and 68.3\%-confidence statistical uncertainty in the denoted parameter as determined by {\tt fitburst}. 

\efRoundThree{The best-fit estimates of all parameters agree with the {\tt simpulse} input values to within reasonable factors of the statistical uncertainties. In order to demonstrate this consistency, we computed ``uncertainty-weighted deviates" of $\tau_r$, i.e., the deviation of each best-fit $\tau_r$ measurement from the mean value of the 100-sample distribution, divided by its corresponding statistical uncertainty. The resultant distribution of uncertainty-weighted $\tau_r$ deviates is shown in Figure \ref{fig:sims}; this distribution appears to reflect a normal distribution, which is expected as the noise of all simulations is derived from a normal distribution. The Anderson-Darling test statistic ($A^2$) for this high-S/N distribution of uncertainty-weighted $\tau_r$ deviates is $A^2 \approx 0.8$, indicating that the null hypothesis -- these deviates are drawn from a normal distribution -- cannot be rejected at a significance level of at least 2.5\%. The uncertainty-weighted deviates for the other {\tt fitburst} parameter yield similar statistics.}

\efRoundThree{Moreover, this consistency between measurement and noise statistics remains valid for a separate set of 100 simulations that use the same simulation parameters as mentioned above, but possess lower S/N values by a factor of $\sim 10$, i.e., S/N $\sim 20$. We obtained best-fit models of these ``low-S/N" bursts with {\tt fitburst} to perform a similar statistical analysis as was done to the high-S/N data. The distribution of uncertainty-weighted $\tau_r$ deviates for this low-S/N data set is also shown in the histogram of Figure \ref{fig:sims}. We found that the low-S/N distribution of uncertainty-weighted $\tau_r$ deviates yielded an Anderson-Darling test statistic $A^2 \approx 0.28$, indicating consistency with a normal distribution. The differences in $A^2$ between the low-S/N and high-S/N data sets is likely not significant due to the low number of samples. We reserve a {\tt fitburst} analysis of bursts with even lower S/N -- on the order of S/N $\sim$ 8--10, corresponding to survey-detection thresholds -- for future work.} 

\efRoundThree{As a final test, we generated 100 simulations of a low-S/N, two-component burst -- i.e., with $N=2$ -- by adding each low-S/N {\tt simpulse} realization described above with a time-delayed copy of itself. The time separation between the two components was large enough to ensure that the signal from each component did not overlap with signal from the other component, but was otherwise arbitrarily chosen. We then analyzed each $N=2$ simulation with {\tt fitburst} in order to find optimal values of the same parameters analyzed above, leading to $N_{\rm fit} = 12$ simultaneously-fit parameters. The uncertainty-weighted deviates for $\tau_r$ from the $N=2$ {\tt fitburst} fits are also shown in the histogram presented in Figure \ref{fig:sims}. The Anderson-Darling test statistic for this $N=2$, low-S/N sample is $A^2 \approx 0.33$, also indicating consistency with a normal distribution despite the substantial increase in {\tt fitburst} degrees of freedom for an $N=2$ model. We are therefore confident that {\tt fitburst} can produce adequate fit statistics for a wide range of burst brightness and degrees of freedom needed for multi-component fits.}

\section{Analysis with CHIME Data} 
\label{sec:analysis}

Using the framework described in Sections \ref{sec:models} and \ref{sec:method}, we generated models with {\tt fitburst} using a series of data products obtained with the CHIME telescope and its backends. We primarily analyzed ``filterbank" (i.e., channelized timeseries) observations collected by the CHIME/FRB instrument \citep{chimefrb18}. The raw CHIME/FRB data comprise $\sim$2 seconds of total-intensity timeseries, with a time resolution of 0.98304 ms, evaluated across the 400-800 MHz bandwidth of the CHIME telescope using 16,384 frequency channels. We extracted dynamic spectra for each FRB by incoherently dedispersing the filterbank using the upon-detection value of DM and $t_0$ determined by the real-time CHIME/FRB detection system, and preserving at least $\pm$80 ms centered on $t_0$. 

\efRoundThree{Prior to fitting, all CHIME/FRB dynamic spectra are normalized, baseline-subtracted, and cleaned of radio frequency interference (RFI) by using a variance-threshold criterion appropriate for CHIME/FRB data.\footnote{These ``pre-processing" algorithms are available as a method within the {\tt DataReader} structure of {\tt fitburst} codebase.} While optional, the application of these or similar external algorithms to uncalibrated data, prior to analysis with {\tt fitburst}, is strongly encouraged in order to minimize the contamination of $D$ by terrestrial RFI signals. Moreover, the definition of $M_{kn}$ and use of Equation \ref{eq:chi2} both presume that off-pulse measurements exhibit ``white" noise that fluctuate about zero flux density; therefore, baseline subtraction of $D$ is implicitly assumed by {\tt fitburst}. Nevertheless, as demonstrated in the following subsections, {\tt fitburst} can adequately model a variety of signal morphology observed in (baseline-subtracted) $D$ even in the presence of minimal RFI. The investigation of {\tt fitburst} performance on $D$ that contain ``red" noise will be the subject of future work.}

For all fits discussed in this work, we set $\nu_r=400.1953125$ MHz\footnote{This value of $\nu_r$ corresponds to the center of the lowest frequency channel digitized by the CHIME telelscope. Moreover, this value is used by the real-time CHIME/FRB detection pipeline in its reporting of DM and $t_0$ values.} and held the \{$\delta=-4$, $\epsilon=-2$\} parameters fixed. All best-fit parameters are reported below with uncertainties determined from the covariance-matrix formalism discussed in Appendix \ref{sec:app_jacobian}; these uncertainties are quoted in parentheses and reflect 68.3\% confidence intervals.

\begin{figure}
    \centering
    \includegraphics[scale=0.3]{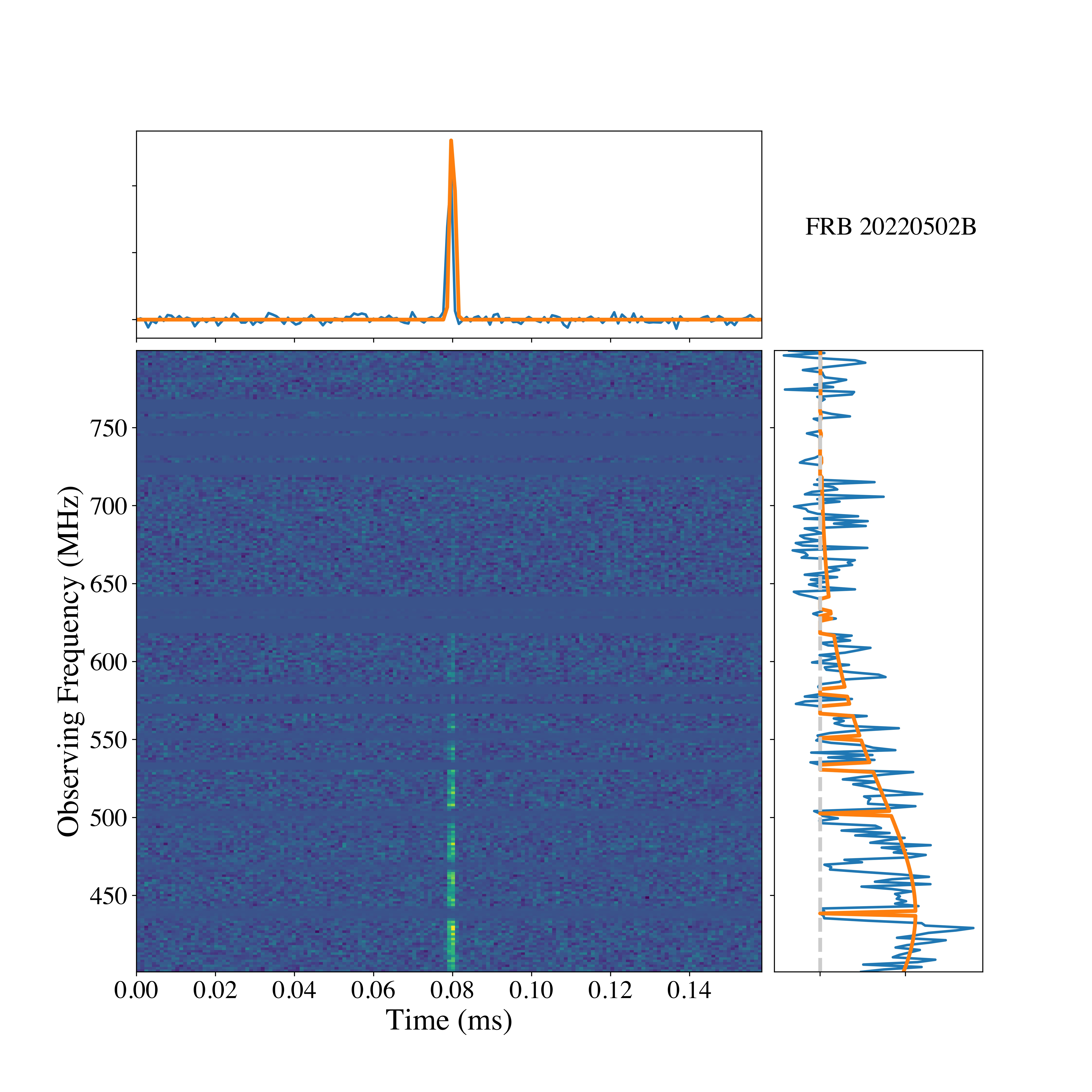}
    \includegraphics[scale=0.3]{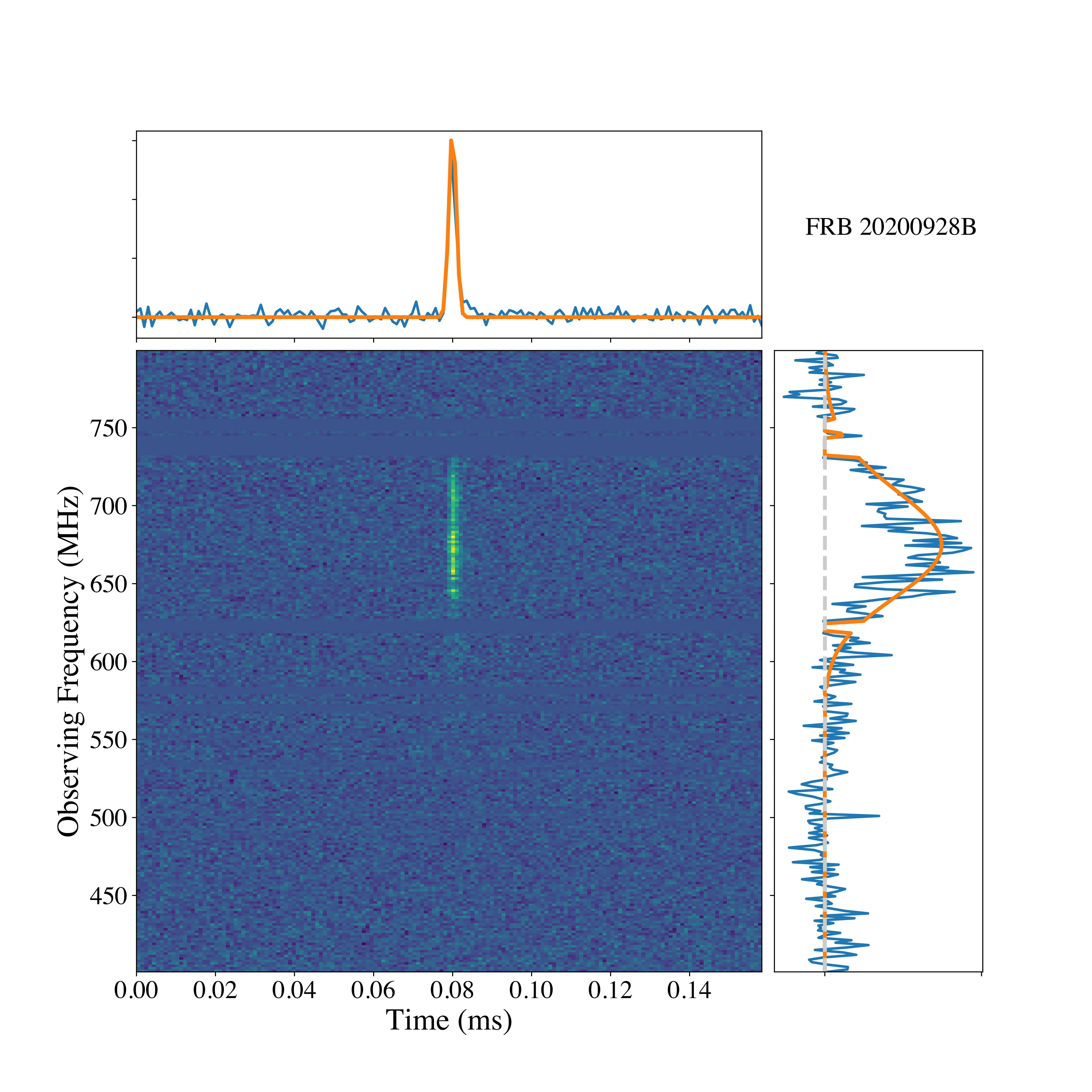}
    \caption{The CHIME/FRB detections of FRBs 20220502B (left) and 20200928B (right), with marginalized timeseries and SEDs shown in the outer panels for each FRB; the orange lines correspond to the best-fit $M_{kn}$ projected onto each axis. The best-fit models for both FRBs were optimized against raw CHIME/FRB data, though the products shown here are downsampled along the frequency axis to 256 channels.}
    \label{fig:data_CHIMEFRB}
\end{figure}

\subsection{Quality of Fits for SEDs}
Figure \ref{fig:data_CHIMEFRB} shows examples of best-fit, single-component dynamic spectra of two CHIME/FRB detections. FRB 20220502B displays a power-law SED while FRB 20200928B exhibits a band-limited, Gaussian-like energy distribution within the CHIME band. For each fit, {\tt fitburst} used the upon-detection estimates of DM and $t_0$ as initial guesses for the least-squares fitting of the model spectrum; all other parameters were initially set to generic values of \{$\alpha=0$, $\beta=0$, $\gamma=0$, $\delta=-4$, $\epsilon=-2$, $\sigma=10$ ms, $\tau_r=0$ ms\}. The \{$\delta$, $\epsilon$, $\tau_r$\} parameters were held fixed to their initial values; all other model parameters were subject to least-squares minimization.

As Figure \ref{fig:data_CHIMEFRB} demonstrates, the {\tt fitburst} RPL model can accommodate a diverse range of SEDs. If both \{$\beta$, $\gamma$\} are simultaneously optimized for FRB 20220502B, then the best-fit {\tt fitburst} model yields moderate values of the amplitude and RPL parameters: $\alpha = -0.583(17)$; $\beta = -16.8(1.1)$; and $\gamma = 3.1(4)$. If instead $\beta=0$ is held fixed for this FRB, then $\alpha=-0.367(11)$ and $\gamma=-3.95(9)$. The case where $\gamma < 0$ conforms with the ``steep-spectrum" SED that is typical of radio pulsars, but the inclusion of $\beta$ as a variable parameter improves the fit by an amount $\Delta\chi^2 \approx 600$ despite only adding one degree of freedom. The significance of $\beta$ characterizes the slight turnover in power at low frequencies that deviates from a purely power-law form.

By contrast, FRB 20200928B yields a best-fit model with substantially different magnitudes of the amplitude and SED parameters: $\alpha = -24.2(1.2)$; $\beta = -198(10)$; and $\gamma = 207(11)$. No model with $\beta=0$ held fixed can adequately describe the band-limited SED for FRB 20200928B, supporting the notion that $\beta$ serves as a measure of Gaussianity as described in Appendix \ref{sec:app_gaussian}. Values with these magnitudes are common for other ``band-limited" signals in the CHIME/FRB data set, especially when the SED is largely encapsulated within the observed bandwidth.

\subsection{Removal of Dispersion Smearing}

\begin{figure}
    \centering
    \includegraphics[scale=0.7]{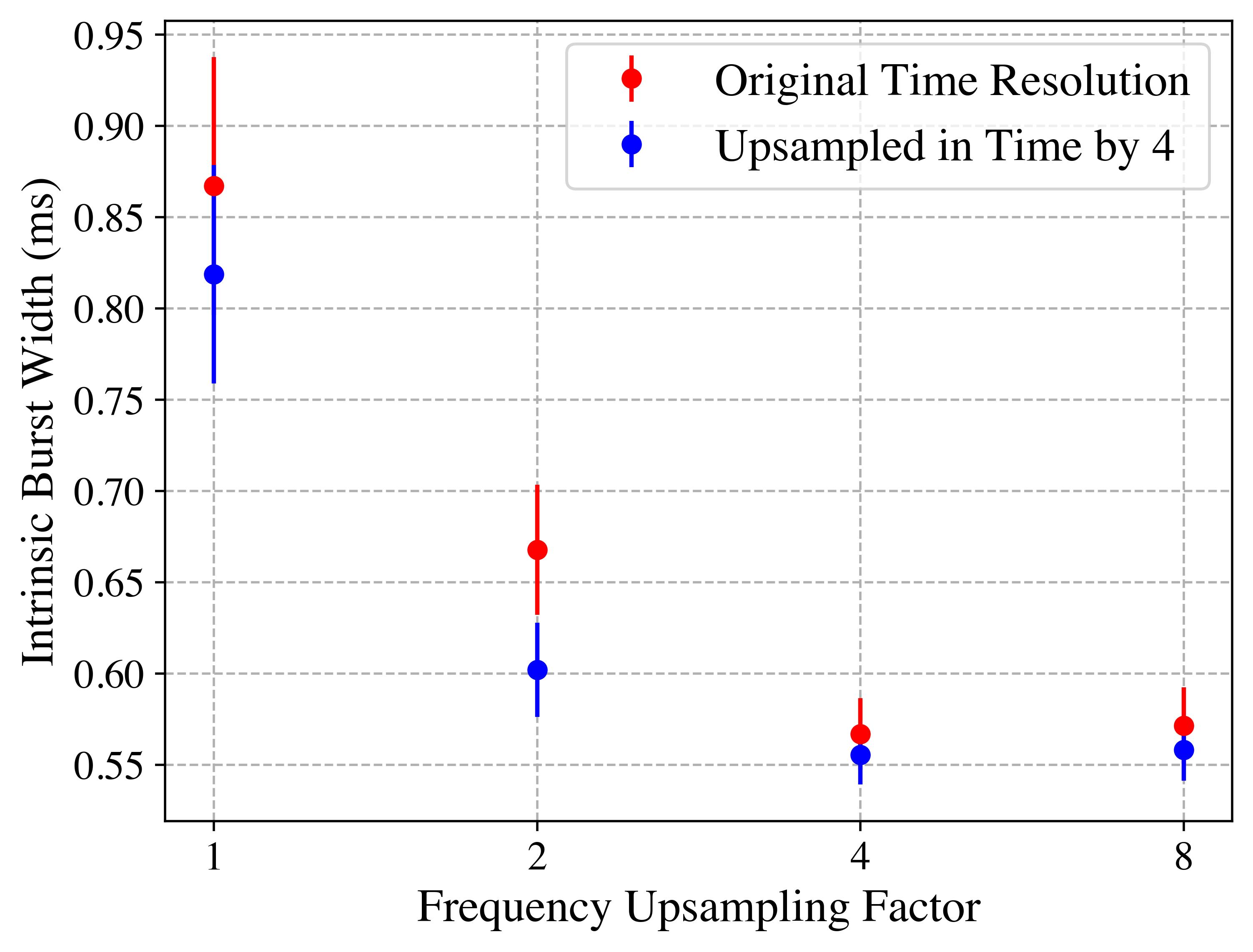}
    \caption{Measurements of $\sigma$ when using different frequency/time upsampling factors to model FRB 20200722B. Each point and uncertainty in this figure corresponds to a best-fit estimate of $\sigma$ made with {\tt fitburst} for a specific amount of time/frequency upsampling of input data while simultaneously optimizing all other morphological parameters, based on the method described in Section \ref{subsec:smearing_method}. The significant differences in $\sigma$ and its statistical uncertainty indicate the presence of dispersion smearing in FRB 20200722B.}
    \label{fig:smearing_104231403}
\end{figure}

As discussed in Section \ref{subsec:smearing_method}, {\tt fitburst} uses an upsampling approach to account for dispersion smearing in channelized, dispersed data. The upsampling factors can be chosen arbitrarily, but should be set to values that are appropriate for the observed signal DM, receiver bandwidth, and time/frequency resolutions of the input data. 

An example of impact in upsampling-factor choices for CHIME/FRB analysis is shown in Figure \ref{fig:smearing_104231403}. The best-fit values and uncertainties of $\sigma$ for FRB 20200722B -- detected by CHIME/FRB with an observed DM of 2,157.572(4) pc cm$^{-3}$ --  clearly decrease for increasing amounts of frequency/time upsampling. This variation demonstrates that dispersion smearing is detectable in the CHIME band at such high DMs. The difference in the two sets of $\sigma$ measurements illustrates the interplay between upsampling-factor choices and resolution of the CHIME/FRB total-intensity data: if there is no upsampling in time, then more upsampling in frequency is needed to better resolve dispersion smearing and thus estimate $\sigma$.

Recent CHIME/FRB results have used {\tt fitburst} to find that intrinsic temporal widths are significantly different between confirmed repeating-FRB sources and those that have not been observed to repeat \citep{chimefrb19c,pgk+21}. This difference has increased in significance as the CHIME/FRB catalog continues to grow \citep{fab+20,chimefrb21,chimefrb23}.

\subsection{Quality of Fits for Scatter-Broadening}
\label{subsec:scattering}

In practice, and unlike the no-scattering model, the scatter-broadened {\tt fitburst} model is sensitive to initial-guess values for \{DM, $t_{0,l}$, $\sigma$, $\tau_r$ \} due to their dependence in the error function. We found that fits for scatter-broadening are aided by following a two-step procedure: first by fitting a model where $\tau_r = 0$ ms is held fixed and all other parameters are optimized; and then fitting a separate, scatter-broadened model where $\tau_r$ is an additional degree of freedom and the initial guess corresponds to the results of the no-scattering model.

\begin{figure}
    \centering
    \includegraphics[scale=0.5]{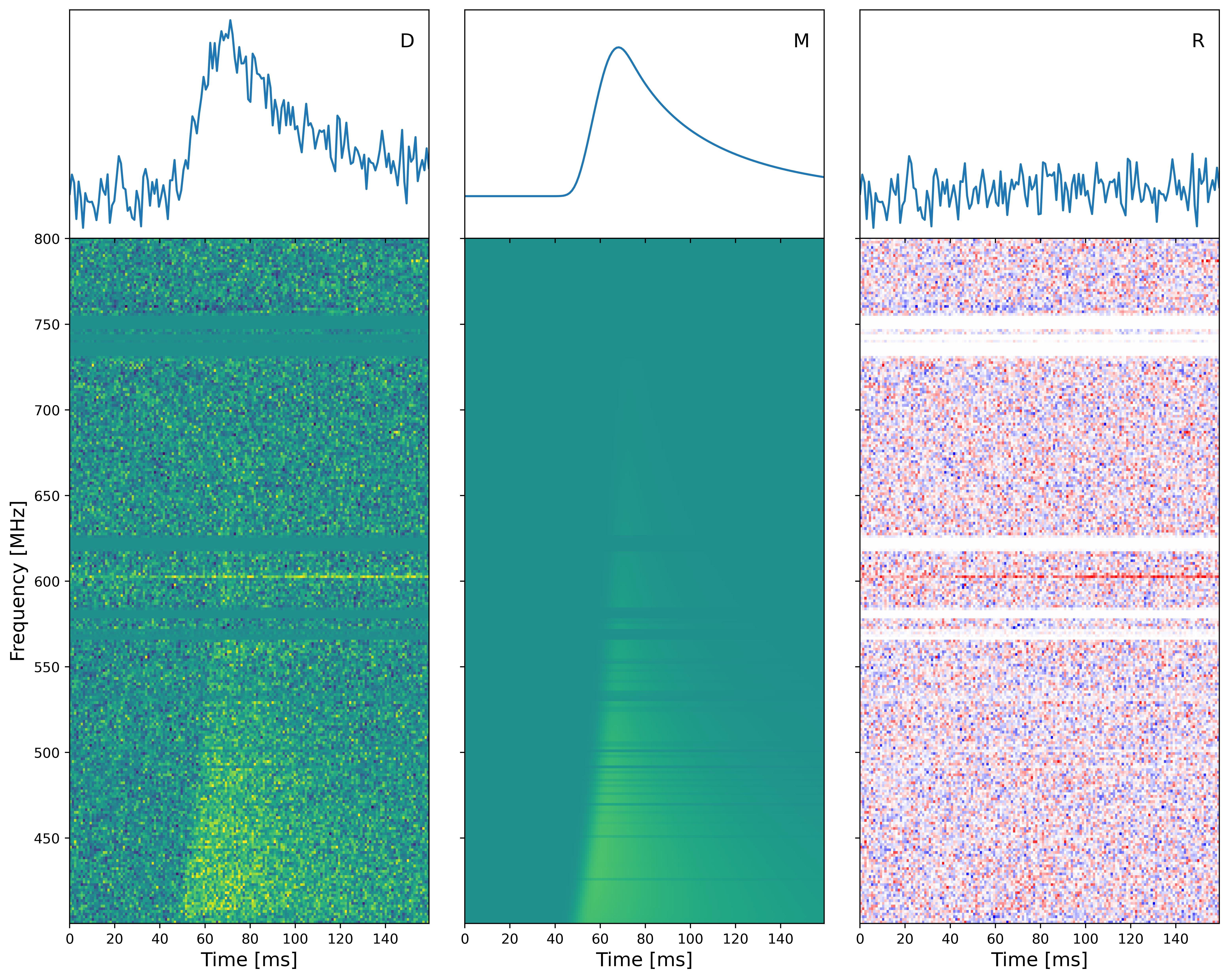}
    \caption{The CHIME/FRB detection of FRB 20210124B (left), optimized scatter-broadened model determined by {\tt fitburst} (middle) and the corresponding residuals (right). The best-fit model was optimized against raw CHIME/FRB data, though each panel shows data that are downsampled along the frequency axis to 256 channels. All parameters except for \{$\delta=-4$, $\epsilon=-2$\} were simultaneously fitted using an initial guess described in Section \ref{subsec:scattering}.}
    \label{fig:scattering_153512030}
\end{figure}

The result of this two-step procedure is shown in Figure \ref{fig:scattering_153512030} for the CHIME/FRB detection of FRB 20210124B. We extracted a dynamic spectrum for FRB 20210124B from CHIME/FRB filterbank data using the DM and $t_0$ values determined from the no-scattering {\tt fitburst} model. We then applied a $N=1$ model where all parameters except \{$\delta=-4$, $\epsilon=-2$\} were simultaneously fitted, yielding a best-fit $\tau_r=\efRoundThree{80}(4)$ ms at $\nu_r=400.1953125$ MHz; the right-most panel of Figure \ref{fig:scattering_153512030} shows the best-fit difference between the model and data, hereafter referred to as ``residuals", which indicates that the model optimally describes the raw data. The inclusion of $\tau_r$ as a fit parameter improves the optimization of the {\tt fitburst} model by an amount $\Delta\chi^2 \approx 700$, despite only adding one degree of freedom between the no-scattering and scatter-broadened models.

Figure \ref{fig:scattering_153512030} also demonstrates the covariance between DM and $\tau_r$, and the important of simultaneously optimizing both parameters. If scatter-broadening is neglected as a model feature, then the scattered signal is absorbed by the no-scattering model as dispersion smearing of the pulse at an incorrect DM. The correct DM is recovered only when incorporating $\tau_r$ as a degree of freedom; for FRB 20210124B, the difference between the optimized no-scattering and scatter-broadened estimates of DM is $\Delta{\rm DM} = 1.16(6)$ pc cm$^{-3}$. 

\subsection{Multi-component Fitting with {\tt fitburst}}
\label{subsec:multicomp}

FRBs often show multi-component structure in their dynamic spectra, typically consisting of distinct bursts that occur at different times and span different frequency ranges. For example, repeating-FRB sources often emit multi-component bursts with ``downward-drifting" substructure, where successive band-limited components occur at decreasing radio frequencies. We therefore designed {\tt fitburst} to accommodate multi-component models in its optimization against input data.

\begin{figure}
    \centering
    \includegraphics[scale=0.5]{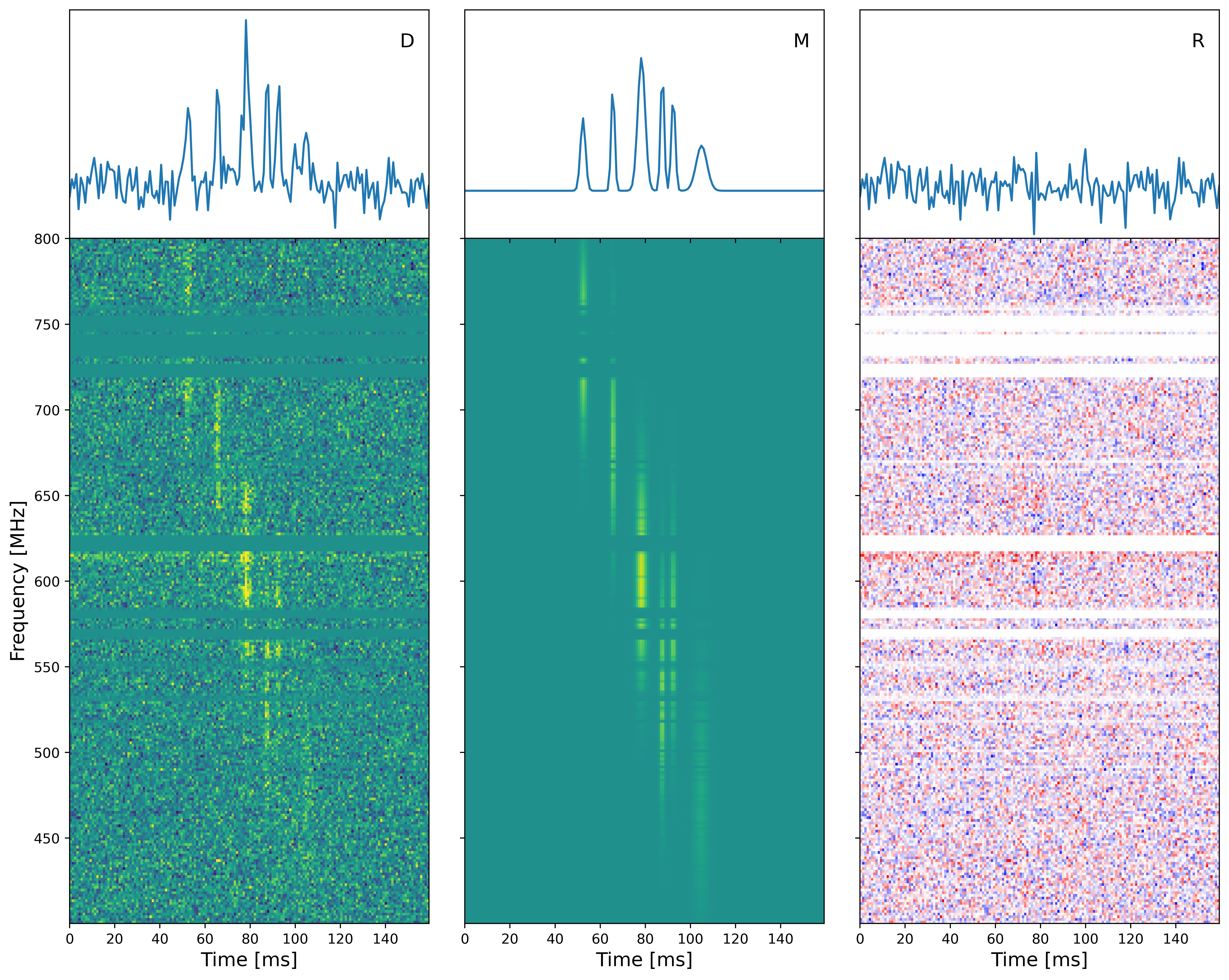}
    \caption{The CHIME/FRB detection of FRB 20211229A (left) and best-fit, $N=6$ model determined by {\tt fitburst}. The best-fit model was optimized against raw CHIME/FRB data, though each panel shows data that are downsampled along the frequency axis to 256 channels. All parameters except for \{$\delta=-4$, $\epsilon=-2$, $\tau_r=0$ ms\} were simultaneously fitted using an initial guess described in Section \ref{subsec:multicomp}.}
    \label{fig:mutlicomp}
\end{figure}

Figure \ref{fig:mutlicomp} shows the data, optimized {\tt fitburst} model and the residuals for FRB 20211229A. For the initial-guess parameters, we used the upon-detection estimate of DM but set each arrival time to the value determined by a peak-finding algorithm applied to the band-averaged timeseries.\footnote{The CHIME/FRB real-time detection pipeline only estimates one representative value of $t_0$ to a single ``event", even if it contains multiple burst components, and so an additional algorithm is necessary for obtaining estimates of $t_{0,l}$. {\tt fitburst} offers such a peak-finding algorithm within its codebase, and its invocation is described in the source-code documentation.} We also arbitrarily set $\nu_r=400.1953125$ MHz for all components. All other per-component parameters were initially set to generic values of \{$\alpha_l=0$, $\beta_l=0$, $\gamma_l=0$, $\sigma_l=0.5$ ms\}. The \{$\delta=-4$, $\epsilon=-2$, $\tau_r=0$ ms\} global parameters were held fixed to their initial values; all other model parameters were subject to least-squares minimization, leading to $N_{\rm fit} = 31$. 

The best-fit model for FRB 20211229A adequately describes each burst component, as indicated by the noise-dominated residuals shown in Figure \ref{fig:mutlicomp}. Moreover, the {\tt fitburst} framework obtained a robust model despite the large number of fit parameters and use of sub-optimal guesses for most quantities. However, the adequacy of the initial guess nonetheless relied on a peak-finding algorithm for establishing the number and arrival-time estimates of burst components. This peak-finding method assumes that the initial guess for DM can sufficiently resolve individual pulses within the band-averaged timeseries; a sufficiently incorrect value of DM will smear pulses and lead to sub-optimal determination of burst components.

\subsection{Fitting to ``Folded" and/or Coherently-Dedispersed Data}
Radio pulsars are often observed using a real-time ``folding" algorithm that averages consecutive pulses into a single representation of improved statistical significance, based on an existing model of neutron-star rotation \citep[e.g.,][]{lk12}. Moreover, modern real-time pulsar instruments usually perform ``coherent dedispersion" \citep{hr75} in order to remove the smearing effects of pulse dispersion at the complex-voltage level. The union of these real-time processing steps form the regular observing mode of the ``CHIME/Pulsar" telescope backend \citep{chimepsr21}, which produces coherently dedispersed timing data for hundreds of pulsars a day, evaluated across 1,024 frequency channels and with time resolution as high as 2.56 $\mu$s.

\begin{figure}
    \centering
    \includegraphics[scale=0.24
    ]{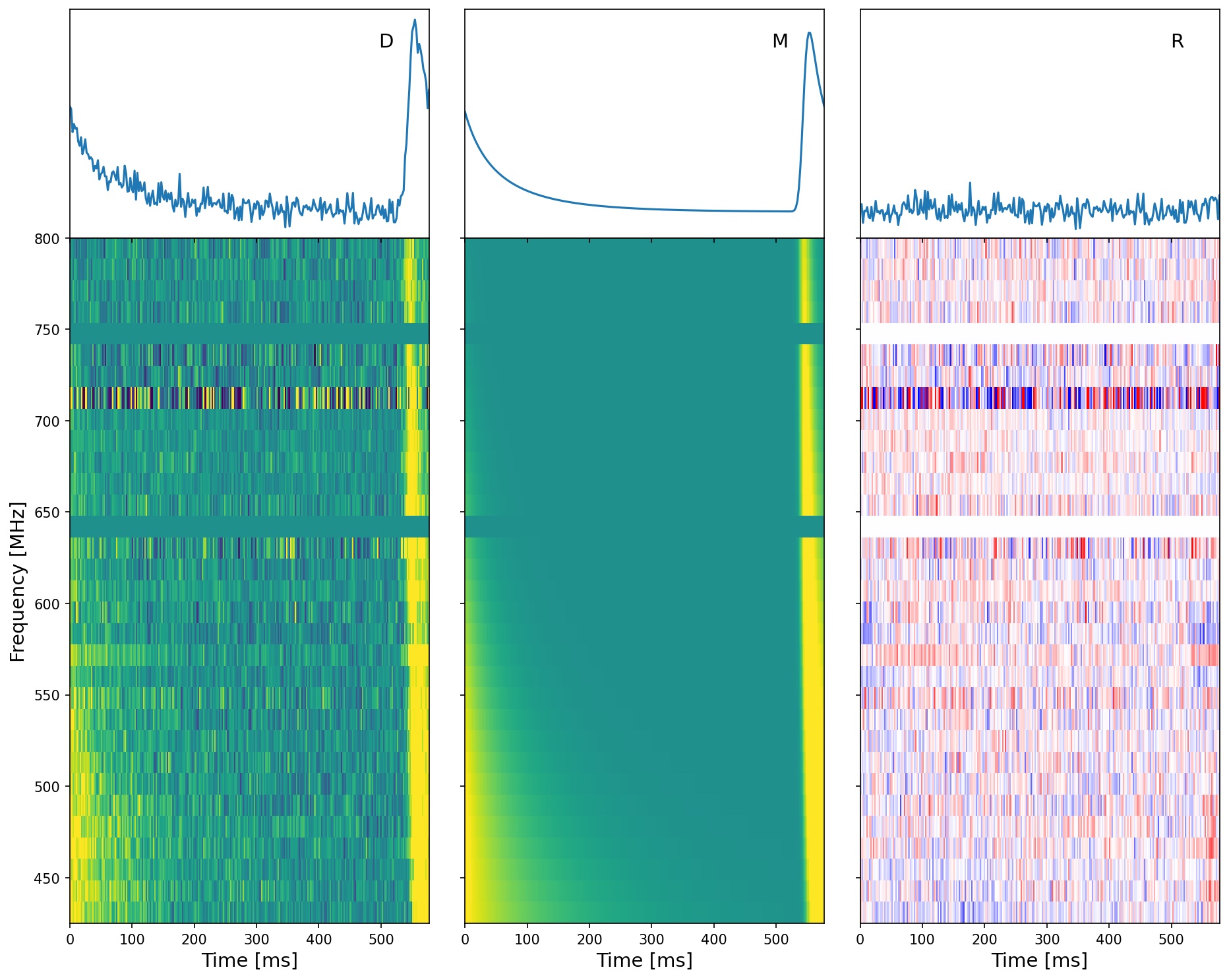}
    \caption{The coherently dedispersed CHIME/Pulsar observation of PSR J2108+4504 on MJD 59438 and best-fit model determined by {\tt fitburst}, whose results were first published by \cite{afm+23}. {\tt fitburst} can adequately account for wrapping of signal in rotational phase, which is important for pulsars with large duty cycle and/or significant scattering-broadening as seen above.}
    \label{fig:snowbirdfit}
\end{figure}

We incorporated features into {\tt fitburst} that enable direct fitting of folded and/or coherently dedispersed versions of $D$ in pulsar data. For coherently-dedispersed data, we constructed {\tt fitburst} to interpret the DM as an offset parameter measured relative to the value used for coherent dedispersion. For folded data, {\tt fitburst} will first generate a realization of $T_{kn,l}$ over a timespan equal to twice the folding period, and then average the two periods into a single realization. An example of these features applied to CHIME/Pulsar timing data is shown in Figure \ref{fig:snowbirdfit} for PSR J2108+4516, a pulsar recently discovered to be orbiting a high-mass OBe star \citep{afm+23}. The two-realization step allows for wrapping of the scattered pulse and thus robust modeling of the folded spectrum.

\subsection{Fitting in the Presence of Scintillation}
\begin{figure}
    \centering
    \includegraphics[scale=0.5]{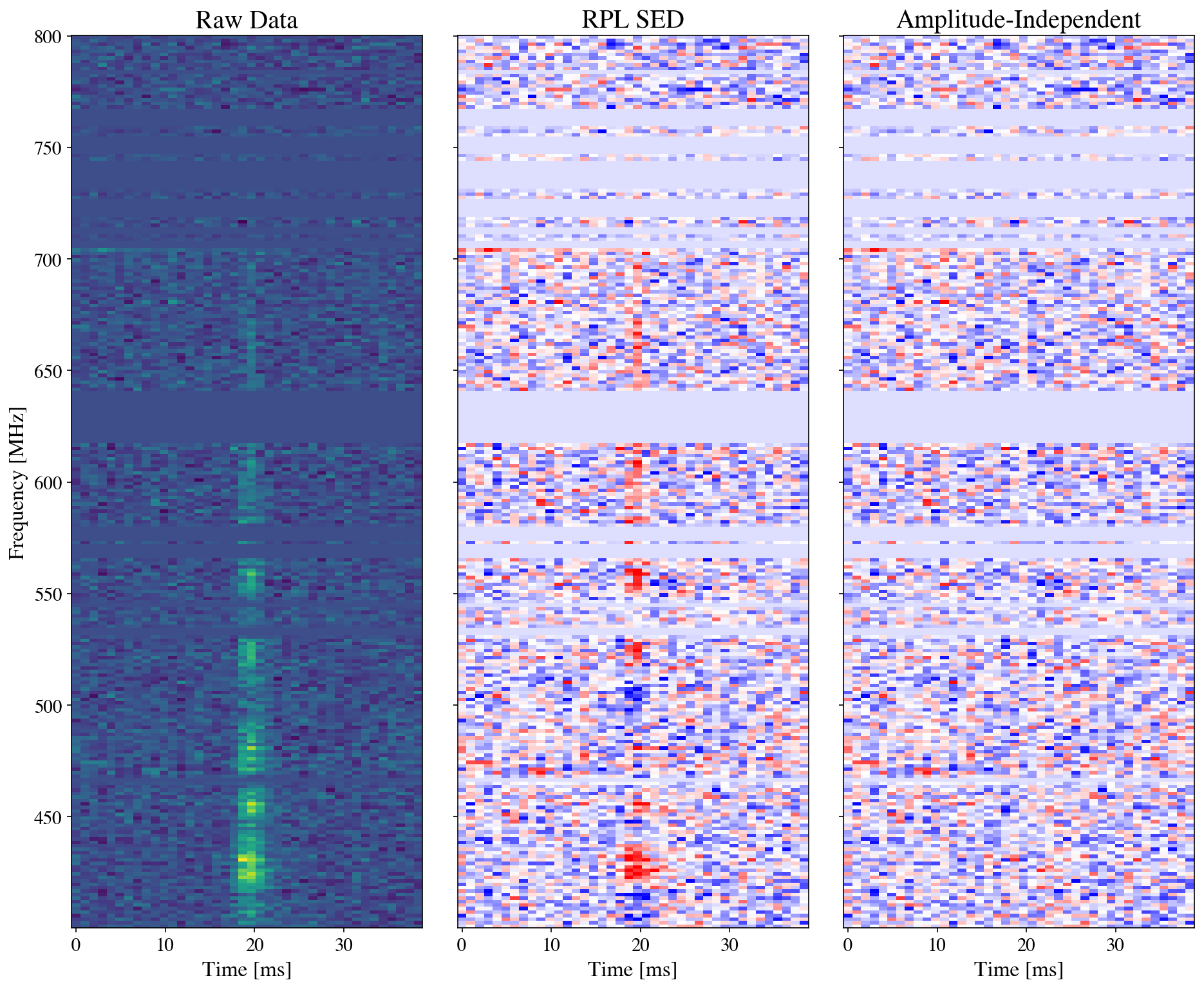}
    \caption{The CHIME/FRB detection of FRB 20230702B (left). The middle panels corresponds to residuals obtained when fitting a ``full", $N=1$ {\tt fitburst} model as described in Section \ref{sec:models}. The right-most panel shows residuals for the same data and model, but computed using the amplitude-independent modeling procedure described in Section \ref{subsec:scintillation}.}
    \label{fig:scintillation}
\end{figure}

A significant number of observed pulsars and FRBs exhibit scintillation in the 400--800-MHz band at which CHIME operates. Figure \ref{fig:scintillation} shows an example of scintillation in FRB 20230702B as detected by CHIME/FRB. A statistically adequate model for $D$ can be obtained when using the RPL SED, though clear residual structure remains. These features do not reflect the smoothly varying decay in intensity from scatter-broadening, and instead indicate incomplete modeling of the SED due to scintillation. These features were nonetheless accounted for when using the amplitude-independent formalism presented in Section \ref{subsec:scintillation}, as shown by the flat residuals in Figure \ref{fig:scintillation}. 

The best-fit values of \{$t_0$, $\sigma$, DM\} between the RFL-SED and amplitude-independent {\tt fitburst} models for FRB 20230702B are statistically consistent, indicating that moderate degrees of scintillation do not negatively impact optimization of $M_{kn}$ if an RPL model of the SED is used. The amplitude-independent optimization of $M_{kn}$ is nonetheless visually superior and indicates that scintillation is present in $D_{kn}$. The per-channel amplitudes afforded by the technique described in Section \ref{subsec:scintillation} can be further interpreted to derive scintillation parameters such as decorrelation bandwidths and timescales. We reserve such efforts to future work. 

\section{Discussion \& Conclusions}
\label{sec:discussion}

The CHIME data set provides an ideal opportunity for developing robust models of dynamic spectra due to its operation at low radio frequencies and apparent diversity in realizations of $D$. However, the framework described in Section \ref{sec:models} is generalized to work with data from any radio telescope. We therefore encourage interested users to apply {\tt fitburst} to their raw data products, in order to obtain a census of morphological parameters for different source classes and at different parts of the electromagnetic spectrum. We conclude this work by discussing various aspects of the {\tt fitburst} framework, both in terms of applications and infrastructure development.

\subsection{Modularity and Runtime of {\tt fitburst}}
{\tt fitburst} is designed to separate distinct operations that, when combined, can contribute to the modeling of $D_{kn}$. This modularity within {\tt fitburst} can therefore be used to simulate realizations of $M_{\rm kn}$ and/or compare existing {\tt fitburst} models without the need for accessing real data or executing a full {\tt fitburst} pipeline. Moreover, different SED models can be incorporated into the {\tt fitburst} framework, so long as exact expressions for the Jacobian-vector and Hessian-matrix components are derived and included as routines within {\tt fitburst}.

The runtime of the {\tt fitburst} optimization algorithm depends on the number of floating point operations needed to estimate Equation \ref{eq:chi2}. This number should largely scale as $N_\nu N_t$ so long as the model-upsampling factors discussed in Section \ref{subsec:smearing_method} are modest in comparison to $N_\nu$ or $N_t$. In order to verify this scaling, we timed the least-squares fitting algorithm within {\tt fitburst} when executed on FRB 20219124B, the CHIME/FRB detection first discussed in Section \ref{subsec:scattering}. We found that the least-squares optimization algorithm needed $\sim 260$ s to optimize all parameters except \{$\delta$, $\epsilon$\} when no downsampling is performed, i.e., when the algorithm is executed on a version of $D_{kn}$ with size ($N_\nu$, $N_t$) = (16384, 162). The runtime decreases linearly when downsampling in frequency by increasing integer multiples of 2, with the algorithm needing only $\sim 3$ s when executed on a version of $D_{kn}$ with size ($N_\nu$, $N_t$) = (265, 162). These runtimes apply for fits that upsample $M_{kn}$ in frequency by a factor of 8 and in time by a factor of 4.

\subsection{Comparison between {\tt fitburst} and {\tt PulsePortraiture}}
The {\tt fitburst} codebase complements {\tt PulsePortaiture} such that the former constructs a synthetic representation of $D$, while the latter assumes knowledge of $D$ (e.g., $\sigma_l$) to extract an accurate estimate of $t_0$ for high-precision pulsar timing. The two frameworks therefore differ in parameterization of $M_{kn}$, with the variable parameters in {\tt PulsePortraiture} consisting only of \{$t_0$, DM, and $\tau_r$\}.\footnote{In addendum, {\tt PulsePortraiture} assigns a single value of arrival time regardless of the number of distinct pulse components observed in $D$. Comparisons of $t_0$ between the two frameworks are therefore non-trivial and we choose to instead focus on the other common parameters.} Moreover, {\tt PulsePortraiture} executes a two-dimensional, template-matching algorithm in the Fourier domain for optimization of $t_0$ while {\tt fitburst} currently optimizes $M_{kn}$ in the time domain.

Despite the differences in frameworks, estimates of \{DM, $\tau_r$\} from {\tt fitburst} can be compared with those determined by {\tt PulsePortraiture} to ensure consistency. Such a comparison was performed by \cite{afm+23} in their pulsar-timing analysis of PSR J2108+4516; this pulsar exhibits persistent and extreme variations in dispersion and scatter-broadening as it orbits its OBe companion star, thus lending itself to simultaneous analysis of morphological and timing properties. The best-fit estimates of $\tau_r$ between {\tt fitburst} and {\tt PulsePortraiture} were found to be statistically consistent. By contrast, the best-fit estimates of DM were found to be globally offset between the two frameworks by $\sim$0.15 pc cm$^{-3}$. Such an overall offset has been noted to arise from measurable differences in the underlying morphology estimated or assumed between the two algorithms \citep[e.g.,][]{pdr14}. The variations in DM for PSR J2108+4516 were nonetheless observed to be the same between {\tt fitburst} and {\tt PulsePortraiture}, indicating that {\tt fitburst} can reliably describe extrinsic effects while simultaneously modeling morphological properties.

\subsection{Comparisons of Best-fit {\tt fitburst} Models}
As described above, there are various scenarios where it is desirable to derive and compare several {\tt fitburst} models for the same $D$. An example was shown in Section \ref{subsec:scattering}, where two models were generated and compared for FRB 20210124B that fitted or ignored morphological parameters that quantify scatter-broadening. In this case, we argued that scatter-broadening was statistically significant since $\Delta\chi^2 = \chi^2_2 - \chi^2_1 \gg \Delta N_{\rm fit} = 1$ between the two models, where $\chi^2_2$ refers to the goodness-of-fit statistic for the model that directly fits for $\tau_r$.

A rigorous comparison between nested models that differ by $\Delta N_{\rm fit} \ge 1$ can be performed using statistical tests. A commonly used scheme is the {\it F}-test, which assumes that the statistic

\begin{equation}
    F = \frac{\Delta\chi^2}{\chi^2_2}\frac{N_{\rm dof,2}}{\Delta N_{\rm dof}}
    \label{eq:Fstat}
\end{equation}

\noindent has an $F$-distribution for parameters \{$\Delta N_{\rm dof}$, $N_{\rm dof,2}$\} under the null hypothesis. We included an initial set of routines within {\tt fitburst} that computes the $p$-value for Equation \ref{eq:Fstat} and uses the cumulative distribution function of the $F$-distribution; when applied to the aforementioned models of FRB 20210124B in Section \ref{subsec:scattering}, we found that the $p$-value $\sim$ $10^{-10}$, indicating that the improvement in fit quality of the scatter-broadened model is unlikely due to chance. Such low $p$-values can be used as an indicator that scatter-broadened models are statistically favored over no-scattering models. We encourage the inclusion of additional tests for discerning statistically superior {\tt fitburst} models.

\begin{figure}
    \centering
    \includegraphics[scale=0.8]{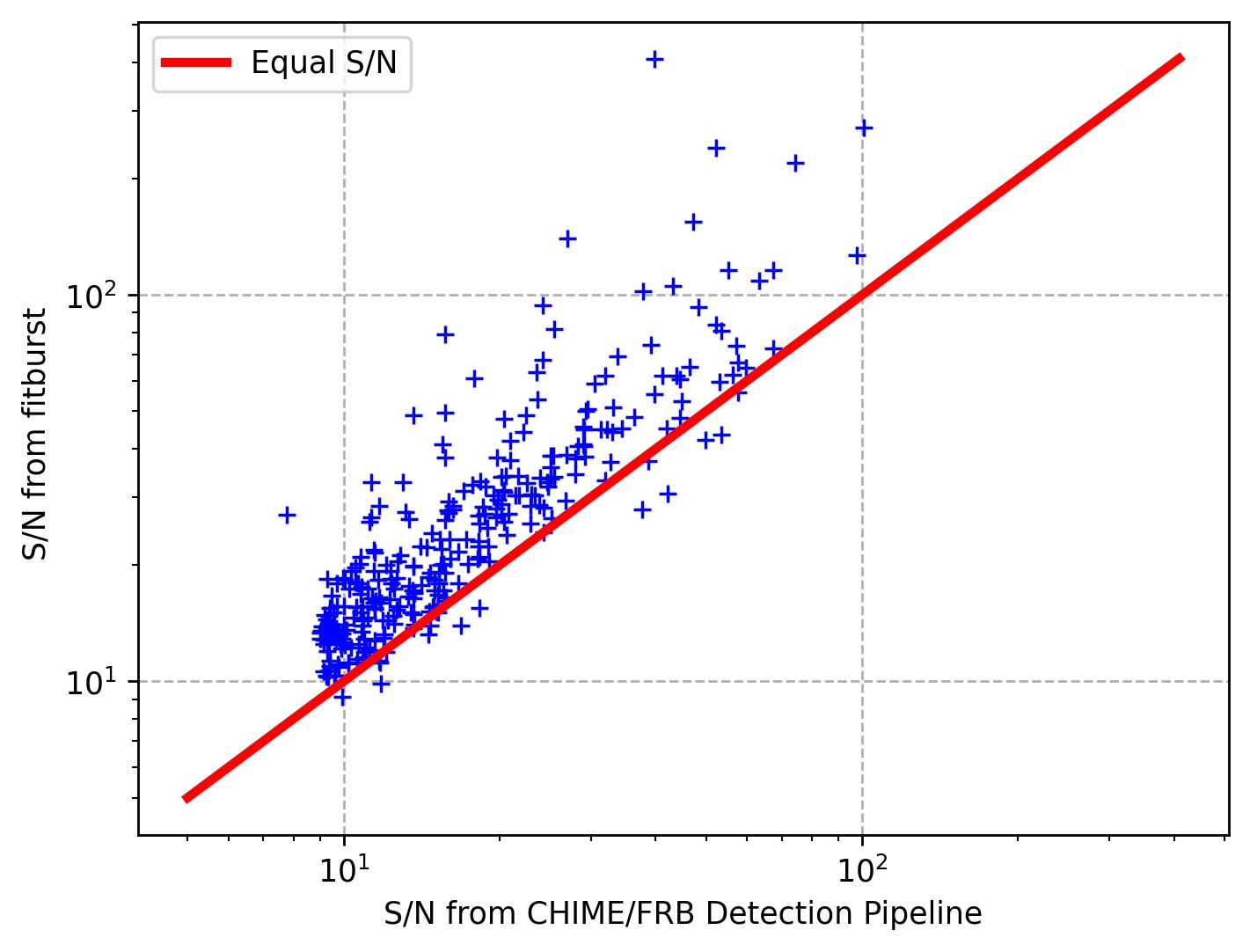}
    \caption{S/N for FRBs published in the first CHIME/FRB catalog, with S/N from best-fit {\tt fitburst} models derived using the method discussed in Section \ref{subsec:snr}. Over 95\% of all {\tt fitburst} models yield S/N greater than the value derived by the real-time CHIME/FRB detection pipeline.}
    \label{fig:snr}
\end{figure}

\subsection{A fit-based Estimate of S/N for {\tt fitburst} Models}
\label{subsec:snr}

A simpler model comparison can be performed by comparing a best-fit {\tt fitburst} model with one where all component amplitudes $A_l = 0$, i.e., a model of zero amplitude. In this case, the value of $\Delta\chi^2$ represents a measure of model significance over the noise in $D$. We therefore defined a measure of signal-to-noise ratio to be S/N = $\sqrt{\Delta \chi^2}$ when comparing best-fit and no-amplitude models. Figure \ref{fig:snr} shows S/N measurements made for signals published in the first CHIME/FRB catalog; all S/N values from {\tt fitburst} are greater or equal to those estimate from the real-time CHIME/FRB detection system, indicating that direct modeling with {\tt fitburst} can extract more information than real-time detection pipelines.

\begin{figure}
    \centering
    \includegraphics[scale=0.7]{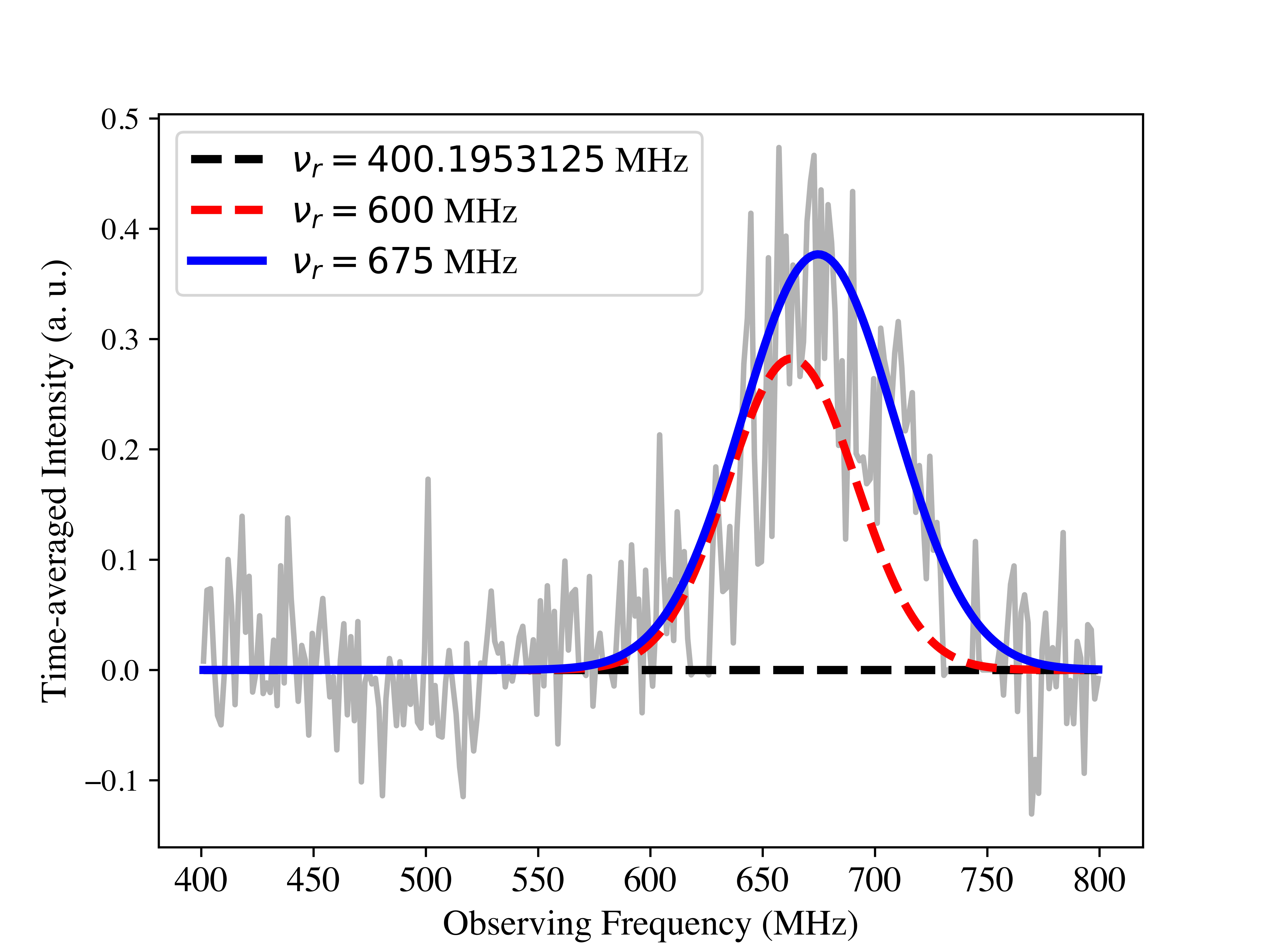}
    \caption{Derived Gaussian SEDs from best-fit {\tt fitburst} models of FRB 20200928B using the RPL SED, where each fit used a different value of $\nu_r$. While all models yielded statistically consistent fits, the choice in $\nu_r$ ultimately impacts the accuracy in translating between RPL and Gaussian parameters for reasons described in Section \ref{subsec:rpl2gauss}.}
    \label{fig:data_Gaussian}
\end{figure}

\subsection{Translating RPL and Gaussian Parameters}
\label{subsec:rpl2gauss}

The analysis presented in Appendix \ref{sec:app_gaussian} presents a framework for connecting the RPL parameters to the parameters of a corresponding Gaussian SED, the central emission frequency $\nu_{\rm c}$ and bandwidth $\sigma_\nu$. Figure \ref{fig:data_Gaussian} displays the same time-integrated spectrum of FRB 20200928B as shown in Figure \ref{fig:data_CHIMEFRB}, along with Gaussian SEDs derived from best-fit spectral and amplitude parameters that each use different fixed values of $\nu_r$. 

While all RPL-based {\tt fitburst} models yield comparable fit statistics, the Gaussian SEDs derived from RPL values clearly vary in accuracy: RPL models with $\nu_r$ closer to the ``true" value of $\nu_c$ produce Gaussian SEDs that better reflect the data. This discrepancy arises due to the approximations made in deriving the relationships presented in Appendix \ref{sec:app_gaussian}, based on the assumption that the band-limited signal is defined over a range of frequencies $\nu/\nu_r \sim 1$; given the low frequencies at which CHIME operates, the condition that $\nu/\nu_r\sim1$ is only met for $\nu_r \approx 675$ MHz, which is closest to the true value. This issue is less prominent for observations over similar bandwidths at higher radio frequencies, where the relative difference between frequency channels is smaller.

\efRoundThree{We note that results from multi-component fitting with {\tt fitburst} can be used to derive statistics that describe spectral behavior seemingly unique to repeating-FRB sources. A notable example is the rate of change in the SED centroid (e.g., $d\nu_{\rm c}/dt$) often observed in mult-component bursts from repeating FRBs, which was first shown to exhibit a negative, linear ramp in FRB 20121102A \citep[e.g.,][]{hss+19}{}. Estimates of the drift rate can be computed from best-fit estimates of the SED parameters and their translation to estimates of \{$\nu_{\rm c}$, $\sigma_{\rm c}$\} at distinct times. For low-frequency observations, we urge caution in first ensuring accuracy of the translation to \{$\nu_{\rm c}$, $\sigma_{\rm c}$\} noted above and in Figure \ref{fig:data_Gaussian}, prior to estimating drift rates from {\tt fitburst} data. More generally, additional care must be taken to account for covariances in all {\tt fitburst} parameters when estimating, e.g., $d\nu_{\rm c}/dt$, as incorrect dedispersion or suboptimal fits of scattering-broadening can dramatically effect constraints on the drift rate \citep[][]{gbp+24}.}

\subsection{Use of Exact vs. Approximate Gradients}
Most modern scientific-computing packages offer optimization routines that numerically estimate $\nabla\chi^2$ through finite-difference methods. While convenient, we found that using the exact form of $\nabla\chi^2$ and the Hessian matrix offers at least three important advantages:

\begin{enumerate}
    \item when executed on CHIME/FRB data, {\tt fitburst} fitting yields a runtime that is $\sim$200\% faster than fits that use the built-in numerical methods of {\tt scipy.optimize.least\_squares};

    \item for a fixed set of initial guesses, {\tt fitburst} is considerably more successful in obtaining a robust model when using the exact form of $\nabla\chi^2$;

    \item statistical uncertainties that are derived from the exact Hessian matrix \efRoundThree{can be} at least $\sim$80\% larger than those derived from the Gauss-Newton approximation of the Hessian, and \efRoundThree{thus provide conservative estimates} of the underlying uncertainties in the fit parameters.
\end{enumerate}

These circumstances are most consequential for efficient characterization of FRB morphology, e.g., for real-time systems like CHIME/FRB that detect several FRBs per day. As next-generation instruments work towards maximizing detection rates and issue near-real-time alerts, {\tt fitburst} can serve as a reliable tool for producing FRB-parameter data sets with robust statistics.

\efRoundThree{In order to confirm the robustness of our methods, we performed a Bayesian analysis of \{DM, $\tau_r$\} for the CHIME/FRB detection of FRB 20210124B that was discussed in Section \ref{subsec:scattering}. We first computed a two-dimensional grid of best-fit $\chi^2$ values with {\tt fitburst} for different, fixed combinations of \{DM, $\tau_r$\} while allowing all other fit parameters to be optimized during each fit; this $\chi^2$ map was then converted to a two-dimensional likelihood density function $p({\rm data}|\{{\rm DM},\tau_r\}) = 0.5\exp(-0.5\Delta\chi^2)$, where $\Delta\chi^2 = \chi^2 - {\rm min}(\chi^2)$. We then used Bayes' theorem and assumed no prior knowledge of either parameter, meaning that the posterior distribution $p(\{{\rm DM},\tau_r\}|{\rm data}) \propto p({\rm data}|\{{\rm DM},\tau_r\})$. The posterior probability distribution function (PDF) for either parameter can then be computed by integrating $p(\{{\rm DM},\tau_r\}|{\rm data})$ over the appropriate coordinate.}

\begin{figure}
    \centering
    \includegraphics[scale=0.7]{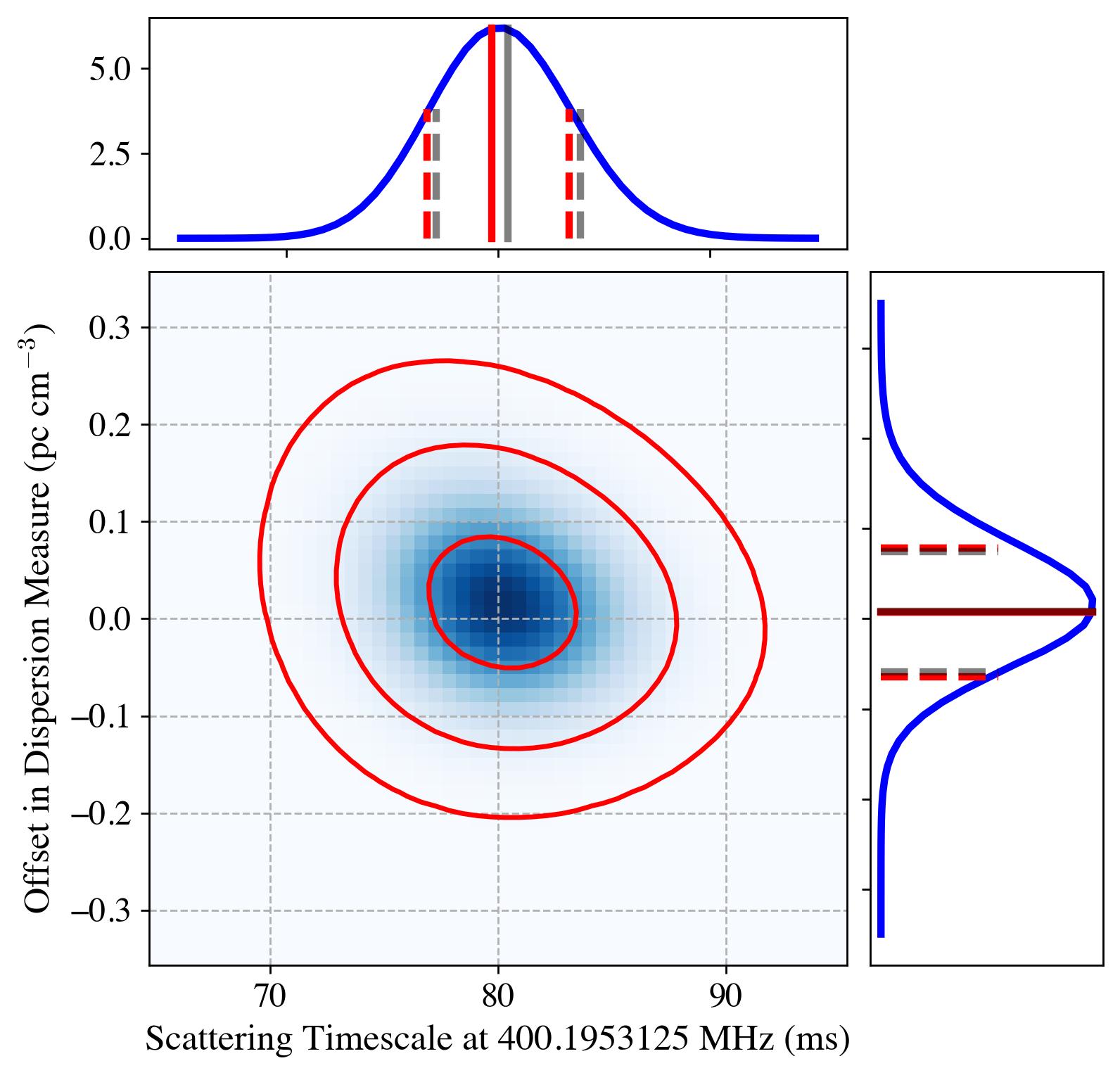}
    \caption{\efRoundThree{The two-dimensional posterior probability density function (blue shading) and marginalized PDFs (blue curves in outer panels) of \{DM, $\tau_r$\} for FRB 20210124B. The inner-to-outermost red contours contain 68.3\%, 95.4\%, and 99.5\% of all probability density, respectively. In the marginalized-PDF panels, the red lines denote the median value (solid) and 68.3\% credible interval (dashed) derived directly from the PDF. The vertical gray lines denote the same measures but instead determined from computation of the exact Hessian matrix as described in Appendix \ref{sec:app_jacobian}. The overlap of these statistics indicates that the exact-Hessian method yields robust measures of uncertainty and covariance.}}
    \label{fig:chisq_grid}
\end{figure}

\efRoundThree{Figure \ref{fig:chisq_grid} presents the results of this Bayesian analysis of \{DM, $\tau_r$\} for FRB 20210124B, including marginalized posterior PDFs for each parameter. Evaluation of the $\chi^2$ grid for FRB 20210124B required $\sim$3 core-days due to the large volume of the CHIME/FRB data set. The median and 68.3\% credible interval for each parameter are essentially identical to the best-fit values and uncertainties computed directly from the exact-Hessian formulation. This consistency confirms that our exact-Hessian method for computing statistical uncertainties produces reasonable estimates.}

\subsection{Future Prospects for {\tt fitburst}}

Several forthcoming works are underway that generate large-scale catalogs of morphological parameters for pulsars and FRBs observed using {\tt fitburst} and CHIME data. One such study comprises the second CHIME/FRB catalog, based on total-intensity filterbank acquired over a 5-yr observing span, that presents measurements for $\sim$4,000 FRBs. Another effort is the first CHIME/FRB catalog of raw-voltage data acquired for $\sim$120 FRBs, which allow for morphological analyses down to timescales as low as 2.56 $\mu$s. A third project consists of a census of single-pulse morphology from several Galactic ``rotating radio transients" observed with the CHIME/Pulsar instrument. A fourth project presents a morphological analysis of bursts from SGR 1935+2154, a Galactic magnetar recently observed to emit pulses with brightnesses comparable to nearby FRB sources \citep[][]{chimefrb20,brb+20}, detected by CHIME/FRB over the past three years \citep{ghi+23}.

As shown above and in prior works, {\tt fitburst} has been developed against CHIME data. However, {\tt fitburst} is generalized to model pulses observed across any part of the radio spectrum, including the lowest frequencies where smearing effects from dispersion are most prominent \citep[e.g.,][]{pck+20,zmb+21,gbp+24}. {\tt fitburst} is modularly designed to enable use of different underlying models of the SED and/or intrinsic shape, though such new feature will require calculations of their partial derivatives for estimation of the covariance matrix.

With this work, we have also made {\tt fitburst} open-source in order to encourage community-based, algorithmic development of features that either improve existing functionality or introduce new methods for characterizing the morphology of radio-transient phenomena. One forthcoming feature will be the incorporation of a Markov Chain Monte Carlo sampling algorithm the employs the {\tt fitburst} model, originally developed by \cite{ghi+23}.


\begin{acknowledgments}    
We acknowledge that CHIME is located on the traditional, ancestral, and unceded territory of the Syilx/Okanagan people. We thank Shami Chatterjee for useful comments on this work. \efRoundThree{We are grateful to the anonymous referee who provided comments that improved the quality of our work.}

We are grateful to the staff of the Dominion Radio Astrophysical Observatory, which is operated by the National Research Council of Canada.  CHIME is funded by a grant from the Canada Foundation for Innovation (CFI) 2012 Leading Edge Fund (Project 31170) and by contributions from the provinces of British Columbia, Qu\'ebec and Ontario. The CHIME/FRB project is funded by a grant from the CFI 2015 Innovation Fund (Project 33213) and by contributions from the provinces of British Columbia and Qu\'ebec, and by the Dunlap Institute for Astronomy and Astrophysics at the University of Toronto. Additional support is provided by the Canadian Institute for Advanced Research (CIFAR), McGill University and the McGill Space Institute thanks to the Trottier Family Foundation, and the University of British Columbia. The CHIME/Pulsar instrument hardware is funded by the Natural Sciences and Engineering Research Council (NSERC) Research Tools and Instruments (RTI-1) grant EQPEQ 458893-2014.
\end{acknowledgments}

\software{
    matplotlib \citep{hun07}, 
    NumPy \citep{hmv+20}
    \psrchive{} \citep{hvm04},
    \pp{} \citep{pdr14},
    SciPy \citep{vgo+20}
}

\appendix

\section{Relationship between Gaussian and Running Power-law Spectral Energy Distributions}
\label{sec:app_gaussian}

As discussed in Section \ref{subsec:rpl}, the running power-law model for SEDs of dynamic spectra can successfully describe both broadband and Gaussian-like FRBs. These two morphological types apparently correspond to distinct values of the ($\beta$, $\gamma$) parameters, with both $\gamma$ and $|\beta| >> 1$ for narrow-band FRBs. In this Appendix, we seek to find the relationship between ($\beta$, $\gamma$) and the standard parameters of a Gaussian profile with mean frequency $\nu_c$ and standard deviation $\sigma_c$, such that

\begin{align}
    \label{eq:rpl_app}
    F(\nu) &= A_{\rm p}(\nu/\nu_r)^{\gamma + \beta\ln(\nu/\nu_r)} = A_{\rm p}e^{(\gamma\ln(\nu/\nu_r) + \beta[\ln(\nu/\nu_r)]^2)} \\
    \label{eq:gauss_app}
    &\approx A_{\rm g}e^{-\frac{1}{2}\frac{(\nu - \nu_c)^2}{\sigma_c^2}}.
\end{align}

The exponents in Equations \ref{eq:rpl_app} and \ref{eq:gauss_app} are incompatible functions of $\nu$ for arbitrary values of $\nu_r$, $\beta$ and $\gamma$. However, each natural logarithm in Equation \ref{eq:gauss_app} can be approximated to yield a parabolic dependence on $\nu$ in regions where $\nu/\nu_r \sim 1$. The Taylor series for $\ln x$ about $x = \nu/\nu_r = 1$ is $\ln x \approx (x-1) - \frac{1}{2}(x-1)^2$; the square of this approximation can be written as $\ln^2x \approx (x-1)^2$ under the assumption that $(x-1) < 1$. Using these approximate relations, the exponent on the right hand side of Equation \ref{eq:rpl_app} can be written as

\begin{equation}
    \label{eq:exp_app}
    \gamma\ln x + \beta[\ln x]^2 \approx \bigg(\beta-\frac{\gamma}{2}\bigg)x^2 + 2(\gamma-\beta)x + \bigg(\beta-\frac{3\gamma}{2}\bigg).
\end{equation}

\noindent We can further rewrite Equation \ref{eq:exp_app} into a binomial form, $(\beta-\gamma/2)(x-h)^2 +k$, by completing the square of its right hand side; this binomial relation is consistent with Equation \ref{eq:exp_app} so long as $h = (\gamma-\beta)/(\gamma/2-\beta)$ and $k = \gamma^2/(\gamma-2\beta)$. Finally, we obtain the relationship between parameters by equating the full expressions for $F(\nu)$ under the above approximation:

\begin{align}  
    F(\nu) &\approx A_{\rm p}e^ke^{(\beta-\gamma/2)[(\nu/\nu_r)-h]^2} \nonumber \\
    &\equiv A_{\rm g}e^{-\frac{1}{2}\frac{(\nu - \nu_c)^2}{\sigma^2}}. \nonumber
\end{align}

\noindent The equivalence between the (approximated) running power-law and Gaussian models requires that

\begin{align}
    \label{eq:amp_relation}
    A_{\rm g} &\approx A_{\rm p}e^{\gamma^2/(\gamma-2\beta)}, \\
    \label{eq:nu_relation}
    \nu_c &\approx \nu_r\frac{(\gamma-\beta)}{(\gamma/2-\beta)}, \textrm{ and} \\
    \label{eq:sigma_relation}
    \sigma_c^2 &\approx -\frac{1}{2}\frac{v_r^2}{\beta - \gamma/2}.
\end{align}

Equations \ref{eq:amp_relation}--\ref{eq:sigma_relation} allow for the conversion from power-law to Gaussian parameters so long as $\nu_r$ is chosen such that the narrow-band signal is defined over $\nu/\nu_r \sim 1$. We can further interpret the $(\beta, \gamma)$ parameters by first inverting Equation \ref{eq:nu_relation} to find $\gamma$ in terms of $\beta$, $\nu_r$ and $\nu_c$, and then inserting this expression in Equation \ref{eq:sigma_relation}; we find that

\begin{align}
    \label{eq:gamma_beta}
    \gamma &= \beta\bigg(\frac{\nu_r-\nu_c}{\nu_r-\nu_c/2}\bigg), \textrm{ and}\\
    \label{eq:sigma_beta}
    \sigma_c^2 &= -\frac{1}{2\beta}\nu_r(\nu_r-\nu_c/2).
\end{align}

\noindent With these relationships, a number of interpretations and predictions can be made:

\begin{itemize}
    \item Equation \ref{eq:sigma_beta} indicates that $\beta$ is a measure of Gaussianity, such that $|\beta| \gg 1$ for values of $\sigma_c \ll$ $\Delta\nu$;
    \item Equation \ref{eq:gamma_beta} indicates that $\gamma$ is a measure of offset from peak emission, with $\gamma\rightarrow0$ as $\nu_r\rightarrow\nu_c$;
    \item $A_{\rm p} \rightarrow A_{\rm g}$ as $\nu_r\rightarrow\nu_c$;
    \item since $\sigma_c>0$, the signs of ($\beta$, $\gamma$) depend on the choice of $\nu_r$; for Catalog 1 of CHIME/FRB, $\nu_r = 400.1953125$ MHz, and so all resolved narrow-band signals in the 400--800 MHz CHIME bandwidth will have $\nu_c>\nu_r$, meaning that $\beta<0$ and $\gamma>0$.
\end{itemize}

\section{The Jacobian and Hessian Matrices for the {\tt fitburst} Model}
\label{sec:app_jacobian}

All least-squares minimization algorithms search for model-parameter values \{$a_i$\} that correspond to the global minimum of Equation \ref{eq:chi2}. In the vicinity of this global minimum, Equation \ref{eq:chi2} can be approximated using a second-order Taylor expansion: 

\begin{equation}
    \label{eq:chi2_expanded}
    \chi^2(\vec{q}) \approx \chi^2(\vec{a}) + \sum_{i}^{N_{\rm fit}}\frac{\partial \chi^2}{\partial q_i}(q_i - a_i) + 
    \frac{1}{2}\sum_{j}^{N_{\rm fit}}\sum_{k}^{N_{\rm fit}}\frac{\partial^2\chi^2}{\partial q_j \partial q_k}(q_j-a_j)(q_k-a_k)
\end{equation}

\noindent where all partial derivatives are to be evaluated at the global minimum, $\vec{q}=\vec{a}$. By definition, all components of the Jacobian matrix $\nabla\chi^2$ are $\partial\chi^2/\partial q_i = 0$ at the global minimum. The collection of mixed partial derivatives in Equation \ref{eq:chi2_expanded} constitute the components of the Hessian matrix, which quantifies the local curvature of the phase space about the global minimum and thus encodes information on the statistical uncertainties.

We determined exact expressions for the Jacobian and Hessian matrices using our definition of model spectrum. For a general fit parameter $q_i$ in a $N$-component model $M_{kn}=\sum_{l}^NM_{kn,l}$, the partial derivatives of $\chi^2$ needed to evaluate Equation \ref{eq:chi2_expanded} are:

\begin{align}
    \label{eq:jacobian}
    \frac{\partial\chi^2}{\partial q_i} &= -2\sum_n^{N_t}\sum_k^{N_\nu}\bigg(\frac{D_{kn}-M_{kn}}{w_k^2}\bigg)\frac{\partial M_{kn}}{\partial q_i} \\
    \label{eq:hessian}
    \frac{\partial^2\chi^2}{\partial q_j \partial q_k} &= 2\sum_n^{N_t}\sum_k^{N_\nu}\frac{1}{w_k^2}\bigg(\frac{\partial M_{kn}}{\partial q_j}\frac{\partial M_{kn}}{\partial q_k}-\big[D_{kn}-M_{kn}\big]\frac{\partial^2M_{kn}}{\partial q_j \partial q_k}\bigg)
\end{align}

\noindent The computation of $\nabla\chi^2$ therefore requires the knowledge of the model's first-derivatives with respect to each fit parameter ($\partial M_{kn}/\partial q_i$), while the Hessian matrix requires knowledge of both first- and mixed-partial derivatives. The use of per-component and global parameters, as discussed in Section \ref{subsec:parameters}, lead to slight differences in the number of terms that contribute to $\partial M_{kn}/\partial q_i$, since global parameters span all models while per-component parameters do not. 

We found the first-order partial derivatives for a scatter-broadened, $N$-component model to be:

\begin{align}
    \label{eq:deriv_amplitude}
    \frac{\partial M_{kn}}{\partial \alpha_l} &= \ln10M_{kn,l} \\
    \label{eq:deriv_sp_running}
    \frac{\partial M_{kn}}{\partial \beta_l} &= \ln^2\bigg(\frac{v_k}{v_r}\bigg)M_{kn, l} \\
    \label{eq:deriv_sp_index}
    \frac{\partial M_{kn}}{\partial \gamma_l} &= \ln\bigg(\frac{v_k}{v_r}\bigg)M_{kn, l} \\
    \label{eq:deriv_burst_width}
    \frac{\partial M_{kn}}{\partial \sigma_l} &= \bigg(\frac{\sigma_l}{\tau_k^2}\bigg)M_{kn, l} + \frac{2A_lF_{k,l}}{\sqrt{\pi}}\bigg(\frac{\partial z_{kn,l}}{\partial \sigma_l}\bigg)\bigg(\frac{\nu_k}{\nu_r}\bigg)^{-\delta}\exp\big[w_{kn,l}\big] \\
    \label{eq:deriv_arrival_time}
    \frac{\partial M_{kn}}{\partial t_{0,l}} &= \bigg(\frac{1}{\tau_k}\bigg)M_{kn,l} + \frac{2A_lF_{k,l}}{\sqrt{\pi}}\bigg(\frac{\partial z_{kn,l}}{\partial t_{0,l}}\bigg)\bigg(\frac{\nu_k}{\nu_r}\bigg)^{-\delta} \exp\big[w_{kn,l}\big] \\
    \label{eq:deriv_sc_time}
    \frac{\partial M_{kn}}{\partial \tau_r} &= \sum_{l}^N\bigg(\frac{1}{\tau_r}\bigg[-\frac{\sigma_l^2}{\tau_k^2} + \frac{(t_{kn}-t_{0,l})}{\tau_k}\bigg]M_{kn,l} + \frac{2A_lF_{k,l}}{\sqrt{\pi}}\bigg(\frac{\partial z_{kn,l}}{\partial\tau_r}\bigg)\bigg(\frac{\nu_k}{\nu_r}\bigg)^{-\delta}\exp\big[w_{kn,l}\big]\bigg) \\
    \label{eq:deriv_sc_index}
    \frac{\partial M_{kn}}{\partial \delta} &= \sum_{l}^N\bigg(-\ln\bigg(\frac{\nu_k}{\nu_r}\bigg)\bigg[1 + \frac{\sigma_l^2}{\tau_k^2} - \frac{[t_{kn}-t_{0,l}]}{\tau_k}\bigg]M_{kn,l} + \frac{2A_lF_{k,l}}{\sqrt{\pi}}\bigg(\frac{\partial z_{kn,l}}{\partial\delta}\bigg)\bigg(\frac{\nu_k}{\nu_r}\bigg)^{-\delta}\exp\big[w_{kn,l}\big]\bigg) \\ 
    \label{eq:deriv_dm_index}
    \frac{\partial M_{kn}}{\partial \epsilon} &= \sum_{l}^N\bigg(\frac{k_{\rm DM}{\rm DM}\big[\nu_k^\epsilon\ln{\nu_k}-\nu_r^\epsilon\ln{\nu_r}\big]}{\tau_k}M_{kn,l} + \frac{2A_lF_{k,l}}{\sqrt{\pi}}\bigg(\frac{\partial z_{kn,l}}{\partial\epsilon}\bigg)\bigg(\frac{\nu_k}{\nu_r}\bigg)^{-\delta}\exp\big[w_{kn,l}\big]\bigg) \\
    \label{eq:deriv_dm}
    \frac{\partial M_{kn}}{\partial {\rm DM}} &= \sum_{l}^N\bigg(\frac{k_{\rm DM}\big[\nu_k^\epsilon-\nu_r^\epsilon\big]}{\tau_k}M_{kn,l} + \frac{2A_lF_{k,l}}{\sqrt{\pi}}\bigg(\frac{\partial z_{kn,l}}{\partial {\rm DM}}\bigg)\bigg(\frac{\nu_k}{\nu_r}\bigg)^{-\delta}\exp\big[w_{kn,l}\big]\bigg)
\end{align}

\noindent where $z_{kn,l}$ is the argument of the error function in Equation \ref{eq:t_def}, and 

\begin{equation}
    w_{kn,l} = \frac{\sigma_l^2}{2\tau_k^2} - \frac{(t_{kn} - t_{0,l})}{\tau_k} - z^2_{kn,l}.
\end{equation}

\noindent In cases where scatter-broadening is insignificant, Equations \ref{eq:deriv_burst_width}--\ref{eq:deriv_dm} are replaced with appropriate derivatives of a Gaussian temporal function since $\tau_r$ and $\delta$ are not defined. 

\begin{figure}
    \centering
    \includegraphics[scale=0.7]{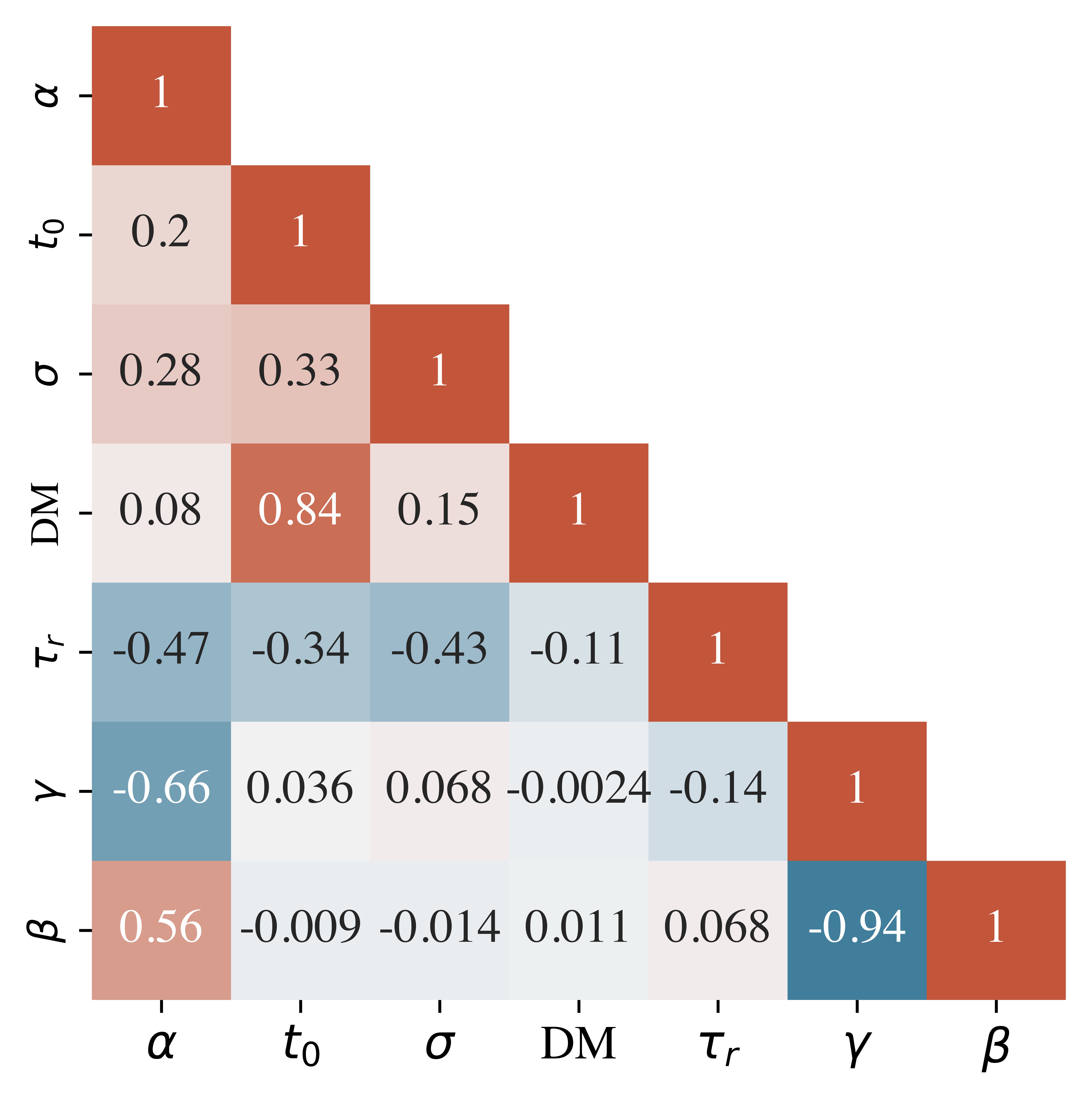}
    \caption{The correlation matrix of the best-fit model for FRB 20210124B (right) presented in Figure \ref{fig:scattering_153512030}.}
    \label{fig:covariance}
\end{figure}

We then computed exact expressions for all possible mixed-partial derivatives $\partial^2\chi^2/\partial q_j\partial q_k$ using Equations \ref{eq:deriv_amplitude}--\ref{eq:deriv_dm}. An important subtlety with the Hessian calculation is that all first derivatives are functions of $M_{kn,l}$, and thus any second derivative with respect to a per-component parameter requires knowledge of $\partial M_{kn,l}/\partial q_l$. The {\tt fitburst} codebase accounts for this step when computing all second derivatives for models with $N > 1$. Once calculated with Equation \ref{eq:hessian}, the Hessian matrix is inverted to yield the covariance matrix. Figure \ref{fig:covariance} shows the best-fit correlation matrix for FRB 20220502B computed using the exact Hessian components provided by {\tt fitburst}.

\bibliographystyle{aasjournal}
\renewcommand{\bibname}{References}

\bibliography{main.bbl}

\begin{thebibliography}{}
\expandafter\ifx\csname natexlab\endcsname\relax\def\natexlab#1{#1}\fi
\providecommand{\url}[1]{\href{#1}{#1}}
\providecommand{\dodoi}[1]{doi:~\href{http://doi.org/#1}{\nolinkurl{#1}}}
\providecommand{\doeprint}[1]{\href{http://ascl.net/#1}{\nolinkurl{http://ascl.net/#1}}}
\providecommand{\doarXiv}[1]{\href{https://arxiv.org/abs/#1}{\nolinkurl{https://arxiv.org/abs/#1}}}

\bibitem[{{Aggarwal} {et~al.}(2021){Aggarwal}, {Agarwal}, {Lewis},
  {Anna-Thomas}, {Tremblay}, {Burke-Spolaor}, {McLaughlin}, \&
  {Lorimer}}]{aal+21}
{Aggarwal}, K., {Agarwal}, D., {Lewis}, E.~F., {et~al.} 2021, \apj, 922, 115,
  \dodoi{10.3847/1538-4357/ac2577}

\bibitem[{{Andersen} {et~al.}(2023){Andersen}, {Fonseca}, {McKee}, {Meyers},
  {Luo}, {Tan}, {Stairs}, {Kaspi}, {van Kerkwijk}, {Bhardwaj}, {Boyle},
  {Crowter}, {Demorest}, {Dong}, {Good}, {Kaczmarek}, {Leung}, {Masui},
  {Naidu}, {Ng}, {Patel}, {Pearlman}, {Pleunis}, {Rafiei-Ravandi}, {Rahman},
  {Ransom}, {Smith}, \& {Tendulkar}}]{afm+23}
{Andersen}, B.~C., {Fonseca}, E., {McKee}, J.~W., {et~al.} 2023, \apj, 943, 57,
  \dodoi{10.3847/1538-4357/aca485}

\bibitem[{{Bochenek} {et~al.}(2020){Bochenek}, {Ravi}, {Belov}, {Hallinan},
  {Kocz}, {Kulkarni}, \& {McKenna}}]{brb+20}
{Bochenek}, C.~D., {Ravi}, V., {Belov}, K.~V., {et~al.} 2020, \nat, 587, 59,
  \dodoi{10.1038/s41586-020-2872-x}

\bibitem[{{CHIME Collaboration} {et~al.}(2022){CHIME Collaboration}, {Amiri},
  {Bandura}, {Boskovic}, {Chen}, {Cliche}, {Deng}, {Denman}, {Dobbs},
  {Fandino}, {Foreman}, {Halpern}, {Hanna}, {Hill}, {Hinshaw}, {H{\"o}fer},
  {Kania}, {Klages}, {Landecker}, {MacEachern}, {Masui}, {Mena-Parra},
  {Milutinovic}, {Mirhosseini}, {Newburgh}, {Nitsche}, {Ordog}, {Pen},
  {Pinsonneault-Marotte}, {Polzin}, {Reda}, {Renard}, {Shaw}, {Siegel},
  {Singh}, {Smegal}, {Tretyakov}, {van Gassen}, {Vanderlinde}, {Wang}, {Wiebe},
  {Willis}, \& {Wulf}}]{chime22}
{CHIME Collaboration}, {Amiri}, M., {Bandura}, K., {et~al.} 2022, \apjs, 261,
  29, \dodoi{10.3847/1538-4365/ac6fd9}

\bibitem[{{CHIME/FRB Collaboration} {et~al.}(2018){CHIME/FRB Collaboration},
  {Amiri}, {Bandura}, {Berger}, {Bhardwaj}, {Boyce}, {Boyle}, {Brar},
  {Burhanpurkar}, {Chawla}, {Chowdhury}, {Cliche}, {Cranmer}, {Cubranic},
  {Deng}, {Denman}, {Dobbs}, {Fandino}, {Fonseca}, {Gaensler}, {Giri},
  {Gilbert}, {Good}, {Guliani}, {Halpern}, {Hinshaw}, {H{\"o}fer}, {Josephy},
  {Kaspi}, {Landecker}, {Lang}, {Liao}, {Masui}, {Mena-Parra}, {Naidu},
  {Newburgh}, {Ng}, {Patel}, {Pen}, {Pinsonneault-Marotte}, {Pleunis}, {Rafiei
  Ravandi}, {Ransom}, {Renard}, {Scholz}, {Sigurdson}, {Siegel}, {Smith},
  {Stairs}, {Tendulkar}, {Vanderlinde}, \& {Wiebe}}]{chimefrb18}
{CHIME/FRB Collaboration}, {Amiri}, M., {Bandura}, K., {et~al.} 2018, \apj,
  863, 48, \dodoi{10.3847/1538-4357/aad188}

\bibitem[{{CHIME/FRB Collaboration} {et~al.}(2019){CHIME/FRB Collaboration},
  {Andersen}, {Bandura}, {Bhardwaj}, {Boubel}, {Boyce}, {Boyle}, {Brar},
  {Cassanelli}, {Chawla}, {Cubranic}, {Deng}, {Dobbs}, {Fandino}, {Fonseca},
  {Gaensler}, {Gilbert}, {Giri}, {Good}, {Halpern}, {Hill}, {Hinshaw},
  {H{\"o}fer}, {Josephy}, {Kaspi}, {Kothes}, {Landecker}, {Lang}, {Li}, {Lin},
  {Masui}, {Mena-Parra}, {Merryfield}, {Mckinven}, {Michilli}, {Milutinovic},
  {Naidu}, {Newburgh}, {Ng}, {Patel}, {Pen}, {Pinsonneault-Marotte}, {Pleunis},
  {Rafiei-Ravandi}, {Rahman}, {Ransom}, {Renard}, {Scholz}, {Siegel}, {Singh},
  {Smith}, {Stairs}, {Tendulkar}, {Tretyakov}, {Vanderlinde}, {Yadav}, \&
  {Zwaniga}}]{chimefrb19c}
{CHIME/FRB Collaboration}, {Andersen}, B.~C., {Bandura}, K., {et~al.} 2019,
  \apjl, 885, L24, \dodoi{10.3847/2041-8213/ab4a80}

\bibitem[{{CHIME/FRB Collaboration} {et~al.}(2020){CHIME/FRB Collaboration},
  {Andersen}, {Bandura}, {Bhardwaj}, {Bij}, {Boyce}, {Boyle}, {Brar},
  {Cassanelli}, {Chawla}, {Chen}, {Cliche}, {Cook}, {Cubranic}, {Curtin},
  {Denman}, {Dobbs}, {Dong}, {Fandino}, {Fonseca}, {Gaensler}, {Giri}, {Good},
  {Halpern}, {Hill}, {Hinshaw}, {H{\"o}fer}, {Josephy}, {Kania}, {Kaspi},
  {Landecker}, {Leung}, {Li}, {Lin}, {Masui}, {McKinven}, {Mena-Parra},
  {Merryfield}, {Meyers}, {Michilli}, {Milutinovic}, {Mirhosseini},
  {M{\"u}nchmeyer}, {Naidu}, {Newburgh}, {Ng}, {Patel}, {Pen},
  {Pinsonneault-Marotte}, {Pleunis}, {Quine}, {Rafiei-Ravandi}, {Rahman},
  {Ransom}, {Renard}, {Sanghavi}, {Scholz}, {Shaw}, {Shin}, {Siegel}, {Singh},
  {Smegal}, {Smith}, {Stairs}, {Tan}, {Tendulkar}, {Tretyakov}, {Vanderlinde},
  {Wang}, {Wulf}, \& {Zwaniga}}]{chimefrb20}
{CHIME/FRB Collaboration}, {Andersen}, B.~C., {Bandura}, K.~M., {et~al.} 2020,
  \nat, 587, 54, \dodoi{10.1038/s41586-020-2863-y}

\bibitem[{{CHIME/FRB Collaboration} {et~al.}(2021){CHIME/FRB Collaboration},
  {Amiri}, {Andersen}, {Bandura}, {Berger}, {Bhardwaj}, {Boyce}, {Boyle},
  {Brar}, {Breitman}, {Cassanelli}, {Chawla}, {Chen}, {Cliche}, {Cook},
  {Cubranic}, {Curtin}, {Deng}, {Dobbs}, {Dong}, {Eadie}, {Fandino}, {Fonseca},
  {Gaensler}, {Giri}, {Good}, {Halpern}, {Hill}, {Hinshaw}, {Josephy},
  {Kaczmarek}, {Kader}, {Kania}, {Kaspi}, {Landecker}, {Lang}, {Leung}, {Li},
  {Lin}, {Masui}, {McKinven}, {Mena-Parra}, {Merryfield}, {Meyers}, {Michilli},
  {Milutinovic}, {Mirhosseini}, {M{\"u}nchmeyer}, {Naidu}, {Newburgh}, {Ng},
  {Patel}, {Pen}, {Petroff}, {Pinsonneault-Marotte}, {Pleunis},
  {Rafiei-Ravandi}, {Rahman}, {Ransom}, {Renard}, {Sanghavi}, {Scholz}, {Shaw},
  {Shin}, {Siegel}, {Sikora}, {Singh}, {Smith}, {Stairs}, {Tan}, {Tendulkar},
  {Vanderlinde}, {Wang}, {Wulf}, \& {Zwaniga}}]{chimefrb21}
{CHIME/FRB Collaboration}, {Amiri}, M., {Andersen}, B.~C., {et~al.} 2021,
  \apjs, 257, 59, \dodoi{10.3847/1538-4365/ac33ab}

\bibitem[{{CHIME/FRB Collaboration} {et~al.}(2023){CHIME/FRB Collaboration},
  {Andersen}, {Bandura}, {Bhardwaj}, {Boyle}, {Brar}, {Cassanelli},
  {Chatterjee}, {Chawla}, {Cook}, {Curtin}, {Dobbs}, {Dong}, {Faber},
  {Fandino}, {Fonseca}, {Gaensler}, {Giri}, {Herrera-Martin}, {Hill}, {Ibik},
  {Josephy}, {Kaczmarek}, {Kader}, {Kaspi}, {Landecker}, {Lanman}, {Lazda},
  {Leung}, {Lin}, {Masui}, {McKinven}, {Mena-Parra}, {Meyers}, {Michilli},
  {Ng}, {Pandhi}, {Pearlman}, {Pen}, {Petroff}, {Pleunis}, {Rafiei-Ravandi},
  {Rahman}, {Ransom}, {Renard}, {Sand}, {Sanghavi}, {Scholz}, {Shah}, {Shin},
  {Siegel}, {Smith}, {Stairs}, {Su}, {Tendulkar}, {Vanderlinde}, {Wang},
  {Wulf}, \& {Zwaniga}}]{chimefrb23}
{CHIME/FRB Collaboration}, {Andersen}, B.~C., {Bandura}, K., {et~al.} 2023,
  \apj, 947, 83, \dodoi{10.3847/1538-4357/acc6c1}

\bibitem[{{CHIME/Pulsar Collaboration} {et~al.}(2021){CHIME/Pulsar
  Collaboration}, {Amiri}, {Bandura}, {Boyle}, {Brar}, {Cliche}, {Crowter},
  {Cubranic}, {Demorest}, {Denman}, {Dobbs}, {Dong}, {Fandino}, {Fonseca},
  {Good}, {Halpern}, {Hill}, {H{\"o}fer}, {Kaspi}, {Landecker}, {Leung}, {Lin},
  {Luo}, {Masui}, {McKee}, {Mena-Parra}, {Meyers}, {Michilli}, {Naidu},
  {Newburgh}, {Ng}, {Patel}, {Pinsonneault-Marotte}, {Ransom}, {Renard},
  {Scholz}, {Shaw}, {Sikora}, {Stairs}, {Tan}, {Tendulkar}, {Tretyakov},
  {Vanderlinde}, {Wang}, \& {Wang}}]{chimepsr21}
{CHIME/Pulsar Collaboration}, {Amiri}, M., {Bandura}, K.~M., {et~al.} 2021,
  \apjs, 255, 5, \dodoi{10.3847/1538-4365/abfdcb}

\bibitem[{{Cordes}(2002)}]{cor02}
{Cordes}, J.~M. 2002, in Astronomical Society of the Pacific Conference Series,
  Vol. 278, Single-Dish Radio Astronomy: Techniques and Applications, ed.
  S.~{Stanimirovic}, D.~{Altschuler}, P.~{Goldsmith}, \& C.~{Salter}, 227--250

\bibitem[{{Cordes} {et~al.}(1986){Cordes}, {Pidwerbetsky}, \&
  {Lovelace}}]{cpl86}
{Cordes}, J.~M., {Pidwerbetsky}, A., \& {Lovelace}, R.~V.~E. 1986, \apj, 310,
  737, \dodoi{10.1086/164728}

\bibitem[{{Fonseca} {et~al.}(2020){Fonseca}, {Andersen}, {Bhardwaj}, {Chawla},
  {Good}, {Josephy}, {Kaspi}, {Masui}, {Mckinven}, {Michilli}, {Pleunis},
  {Shin}, {Tendulkar}, {Bandura}, {Boyle}, {Brar}, {Cassanelli}, {Cubranic},
  {Dobbs}, {Dong}, {Gaensler}, {Hinshaw}, {Landecker}, {Leung}, {Li}, {Lin},
  {Mena-Parra}, {Merryfield}, {Naidu}, {Ng}, {Patel}, {Pen}, {Rafiei-Ravandi},
  {Rahman}, {Ransom}, {Scholz}, {Smith}, {Stairs}, {Vanderlinde}, {Yadav}, \&
  {Zwaniga}}]{fab+20}
{Fonseca}, E., {Andersen}, B.~C., {Bhardwaj}, M., {et~al.} 2020, \apjl, 891,
  L6, \dodoi{10.3847/2041-8213/ab7208}

\bibitem[{{Geyer} \& {Karastergiou}(2016)}]{gk16}
{Geyer}, M., \& {Karastergiou}, A. 2016, \mnras, 462, 2587,
  \dodoi{10.1093/mnras/stw1724}

\bibitem[{{Giri} {et~al.}(2023){Giri}, {Andersen}, {Chawla}, {Curtin},
  {Fonseca}, {Kaspi}, {Lin}, {Masui}, {Sand}, {Scholz}, {Abbott}, {Dong},
  {Gaensler}, {Leung}, {Michilli}, {Bhardwaj}, {M{\"u}nchmeyer}, {Pandhi},
  {Pearlman}, {Pleunis}, {Rafiei-Ravandi}, {Reda}, {Shin}, {Smith}, {Stairs},
  {Stenning}, \& {Tendulkar}}]{ghi+23}
{Giri}, U., {Andersen}, B.~C., {Chawla}, P., {et~al.} 2023, arXiv e-prints,
  arXiv:2310.16932, \dodoi{10.48550/arXiv.2310.16932}

\bibitem[{{Gopinath} {et~al.}(2024){Gopinath}, {Bassa}, {Pleunis}, {Hessels},
  {Chawla}, {Keane}, {Kondratiev}, {Michilli}, \& {Nimmo}}]{gbp+24}
{Gopinath}, A., {Bassa}, C.~G., {Pleunis}, Z., {et~al.} 2024, \mnras, 527,
  9872, \dodoi{10.1093/mnras/stad3856}

\bibitem[{{Hankins} \& {Rickett}(1975)}]{hr75}
{Hankins}, T.~H., \& {Rickett}, B.~J. 1975, Methods in Computational Physics,
  14, 55, \dodoi{10.1016/B978-0-12-460814-6.50007-3}

\bibitem[{{Harris} {et~al.}(2020){Harris}, {Millman}, {van der Walt},
  {Gommers}, {Virtanen}, {Cournapeau}, {Wieser}, {Taylor}, {Berg}, {Smith},
  {Kern}, {Picus}, {Hoyer}, {van Kerkwijk}, {Brett}, {Haldane},
  {Fern{\'{a}}ndez del R{\'{i}}o}, {Wiebe}, {Peterson},
  {G{\'{e}}rard-Marchant}, {Sheppard}, {Reddy}, {Weckesser}, {Abbasi},
  {Gohlke}, \& {Oliphant}}]{hmv+20}
{Harris}, C.~R., {Millman}, K.~J., {van der Walt}, S.~J., {et~al.} 2020,
  Nature, 585, 357, \dodoi{10.1038/s41586-020-2649-2}

\bibitem[{{Hessels} {et~al.}(2019){Hessels}, {Spitler}, {Seymour}, {Cordes},
  {Michilli}, {Lynch}, {Gourdji}, {Archibald}, {Bassa}, {Bower}, {Chatterjee},
  {Connor}, {Crawford}, {Deneva}, {Gajjar}, {Kaspi}, {Keimpema}, {Law},
  {Marcote}, {McLaughlin}, {Paragi}, {Petroff}, {Ransom}, {Scholz}, {Stappers},
  \& {Tendulkar}}]{hss+19}
{Hessels}, J.~W.~T., {Spitler}, L.~G., {Seymour}, A.~D., {et~al.} 2019, \apjl,
  876, L23, \dodoi{10.3847/2041-8213/ab13ae}

\bibitem[{{Hotan} {et~al.}(2004){Hotan}, {van Straten}, \&
  {Manchester}}]{hvm04}
{Hotan}, A.~W., {van Straten}, W., \& {Manchester}, R.~N. 2004, \pasa, 21, 302,
  \dodoi{10.1071/AS04022}

\bibitem[{Hunter(2007)}]{hun07}
Hunter, J.~D. 2007, Computing in Science \& Engineering, 9, 90,
  \dodoi{10.1109/MCSE.2007.55}

\bibitem[{{Jankowski}(2022)}]{jan22}
{Jankowski}, F. 2022, {Scatfit: Scattering fits of time domain radio signals
  (Fast Radio Bursts or pulsars)}, Astrophysics Source Code Library, record
  ascl:2208.003.
\newblock \doeprint{2208.003}

\bibitem[{{Kulkarni}(2020)}]{kul20}
{Kulkarni}, S.~R. 2020, arXiv e-prints, arXiv:2007.02886,
  \dodoi{10.48550/arXiv.2007.02886}

\bibitem[{{Lentati} {et~al.}(2017){Lentati}, {Kerr}, {Dai}, {Hobson},
  {Shannon}, {Hobbs}, {Bailes}, {Bhat}, {Burke-Spolaor}, {Coles}, {Dempsey},
  {Lasky}, {Levin}, {Manchester}, {Os{\l}owski}, {Ravi}, {Reardon}, {Rosado},
  {Spiewak}, {van Straten}, {Toomey}, {Wang}, {Wen}, {You}, \& {Zhu}}]{lkd+17}
{Lentati}, L., {Kerr}, M., {Dai}, S., {et~al.} 2017, \mnras, 466, 3706,
  \dodoi{10.1093/mnras/stw3359}

\bibitem[{{Liu} {et~al.}(2014){Liu}, {Desvignes}, {Cognard}, {Stappers},
  {Verbiest}, {Lee}, {Champion}, {Kramer}, {Freire}, \& {Karuppusamy}}]{ldc+14}
{Liu}, K., {Desvignes}, G., {Cognard}, I., {et~al.} 2014, \mnras, 443, 3752,
  \dodoi{10.1093/mnras/stu1420}

\bibitem[{{Lorimer} \& {Kramer}(2012)}]{lk12}
{Lorimer}, D.~R., \& {Kramer}, M. 2012, {Handbook of Pulsar Astronomy}
  ({Cambridge University Press})

\bibitem[{{Masui} {et~al.}(2015){Masui}, {Lin}, {Sievers}, {Anderson}, {Chang},
  {Chen}, {Ganguly}, {Jarvis}, {Kuo}, {Li}, {Liao}, {McLaughlin}, {Pen},
  {Peterson}, {Roman}, {Timbie}, {Voytek}, \& {Yadav}}]{mls+15}
{Masui}, K.~W., {Lin}, H.-H., {Sievers}, J., {et~al.} 2015, \nat, 528, 523,
  \dodoi{10.1038/nature15769}

\bibitem[{{McKinnon}(2014)}]{mck14}
{McKinnon}, M.~M. 2014, \pasp, 126, 476, \dodoi{10.1086/676975}

\bibitem[{{Merryfield} {et~al.}(2023){Merryfield}, {Tendulkar}, {Shin},
  {Andersen}, {Josephy}, {Good}, {Dong}, {Masui}, {Lang}, {M{\"u}nchmeyer},
  {Brar}, {Cassanelli}, {Dobbs}, {Fonseca}, {Kaspi}, {Mena-Parra}, {Pleunis},
  {Rafiei-Ravandi}, {Sand}, {Scholz}, {Smith}, \& {Stairs}}]{mts+23}
{Merryfield}, M., {Tendulkar}, S.~P., {Shin}, K., {et~al.} 2023, \aj, 165, 152,
  \dodoi{10.3847/1538-3881/ac9ab5}

\bibitem[{{Parent} {et~al.}(2020){Parent}, {Chawla}, {Kaspi}, {Agazie},
  {Blumer}, {DeCesar}, {Fiore}, {Fonseca}, {Hessels}, {Kaplan}, {Kondratiev},
  {LaRose}, {Levin}, {Lewis}, {Lynch}, {McEwen}, {McLaughlin}, {Mingyar}, {Al
  Noori}, {Ransom}, {Roberts}, {Schmiedekamp}, {Schmiedekamp}, {Siemens},
  {Spiewak}, {Stairs}, {Surnis}, {Swiggum}, \& {van Leeuwen}}]{pck+20}
{Parent}, E., {Chawla}, P., {Kaspi}, V.~M., {et~al.} 2020, \apj, 904, 92,
  \dodoi{10.3847/1538-4357/abbdf6}

\bibitem[{{Pennucci}(2019)}]{pen19}
{Pennucci}, T.~T. 2019, \apj, 871, 34, \dodoi{10.3847/1538-4357/aaf6ef}

\bibitem[{{Pennucci} {et~al.}(2014){Pennucci}, {Demorest}, \& {Ransom}}]{pdr14}
{Pennucci}, T.~T., {Demorest}, P.~B., \& {Ransom}, S.~M. 2014, \apj, 790, 93,
  \dodoi{10.1088/0004-637X/790/2/93}

\bibitem[{{Planck Collaboration} {et~al.}(2020){Planck Collaboration},
  {Aghanim}, {Akrami}, {Ashdown}, {Aumont}, {Baccigalupi}, {Ballardini},
  {Banday}, {Barreiro}, {Bartolo}, {Basak}, {Battye}, {Benabed}, {Bernard},
  {Bersanelli}, {Bielewicz}, {Bock}, {Bond}, {Borrill}, {Bouchet}, {Boulanger},
  {Bucher}, {Burigana}, {Butler}, {Calabrese}, {Cardoso}, {Carron},
  {Challinor}, {Chiang}, {Chluba}, {Colombo}, {Combet}, {Contreras}, {Crill},
  {Cuttaia}, {de Bernardis}, {de Zotti}, {Delabrouille}, {Delouis}, {Di
  Valentino}, {Diego}, {Dor{\'e}}, {Douspis}, {Ducout}, {Dupac}, {Dusini},
  {Efstathiou}, {Elsner}, {En{\ss}lin}, {Eriksen}, {Fantaye}, {Farhang},
  {Fergusson}, {Fernandez-Cobos}, {Finelli}, {Forastieri}, {Frailis},
  {Fraisse}, {Franceschi}, {Frolov}, {Galeotta}, {Galli}, {Ganga},
  {G{\'e}nova-Santos}, {Gerbino}, {Ghosh}, {Gonz{\'a}lez-Nuevo}, {G{\'o}rski},
  {Gratton}, {Gruppuso}, {Gudmundsson}, {Hamann}, {Handley}, {Hansen},
  {Herranz}, {Hildebrandt}, {Hivon}, {Huang}, {Jaffe}, {Jones}, {Karakci},
  {Keih{\"a}nen}, {Keskitalo}, {Kiiveri}, {Kim}, {Kisner}, {Knox},
  {Krachmalnicoff}, {Kunz}, {Kurki-Suonio}, {Lagache}, {Lamarre}, {Lasenby},
  {Lattanzi}, {Lawrence}, {Le Jeune}, {Lemos}, {Lesgourgues}, {Levrier},
  {Lewis}, {Liguori}, {Lilje}, {Lilley}, {Lindholm}, {L{\'o}pez-Caniego},
  {Lubin}, {Ma}, {Mac{\'\i}as-P{\'e}rez}, {Maggio}, {Maino}, {Mandolesi},
  {Mangilli}, {Marcos-Caballero}, {Maris}, {Martin}, {Martinelli},
  {Mart{\'\i}nez-Gonz{\'a}lez}, {Matarrese}, {Mauri}, {McEwen}, {Meinhold},
  {Melchiorri}, {Mennella}, {Migliaccio}, {Millea}, {Mitra},
  {Miville-Desch{\^e}nes}, {Molinari}, {Montier}, {Morgante}, {Moss}, {Natoli},
  {N{\o}rgaard-Nielsen}, {Pagano}, {Paoletti}, {Partridge}, {Patanchon},
  {Peiris}, {Perrotta}, {Pettorino}, {Piacentini}, {Polastri}, {Polenta},
  {Puget}, {Rachen}, {Reinecke}, {Remazeilles}, {Renzi}, {Rocha}, {Rosset},
  {Roudier}, {Rubi{\~n}o-Mart{\'\i}n}, {Ruiz-Granados}, {Salvati}, {Sandri},
  {Savelainen}, {Scott}, {Shellard}, {Sirignano}, {Sirri}, {Spencer},
  {Sunyaev}, {Suur-Uski}, {Tauber}, {Tavagnacco}, {Tenti}, {Toffolatti},
  {Tomasi}, {Trombetti}, {Valenziano}, {Valiviita}, {Van Tent}, {Vibert},
  {Vielva}, {Villa}, {Vittorio}, {Wandelt}, {Wehus}, {White}, {White},
  {Zacchei}, \& {Zonca}}]{planck20}
{Planck Collaboration}, {Aghanim}, N., {Akrami}, Y., {et~al.} 2020, \aap, 641,
  A6, \dodoi{10.1051/0004-6361/201833910}

\bibitem[{{Pleunis} {et~al.}(2021{\natexlab{a}}){Pleunis}, {Good}, {Kaspi},
  {Mckinven}, {Ransom}, {Scholz}, {Bandura}, {Bhardwaj}, {Boyle}, {Brar},
  {Cassanelli}, {Chawla}, {Dong}, {Fonseca}, {Gaensler}, {Josephy},
  {Kaczmarek}, {Leung}, {Lin}, {Masui}, {Mena-Parra}, {Michilli}, {Ng},
  {Patel}, {Rafiei-Ravandi}, {Rahman}, {Sanghavi}, {Shin}, {Smith}, {Stairs},
  \& {Tendulkar}}]{pgk+21}
{Pleunis}, Z., {Good}, D.~C., {Kaspi}, V.~M., {et~al.} 2021{\natexlab{a}},
  \apj, 923, 1, \dodoi{10.3847/1538-4357/ac33ac}

\bibitem[{{Pleunis} {et~al.}(2021{\natexlab{b}}){Pleunis}, {Michilli}, {Bassa},
  {Hessels}, {Naidu}, {Andersen}, {Chawla}, {Fonseca}, {Gopinath}, {Kaspi},
  {Kondratiev}, {Li}, {Bhardwaj}, {Boyle}, {Brar}, {Cassanelli}, {Gupta},
  {Josephy}, {Karuppusamy}, {Keimpema}, {Kirsten}, {Leung}, {Marcote}, {Masui},
  {Mckinven}, {Meyers}, {Ng}, {Nimmo}, {Paragi}, {Rahman}, {Scholz}, {Shin},
  {Smith}, {Stairs}, \& {Tendulkar}}]{zmb+21}
{Pleunis}, Z., {Michilli}, D., {Bassa}, C.~G., {et~al.} 2021{\natexlab{b}},
  \apjl, 911, L3, \dodoi{10.3847/2041-8213/abec72}

\bibitem[{{Ravi}(2019)}]{rav19}
{Ravi}, V. 2019, \mnras, 482, 1966, \dodoi{10.1093/mnras/sty1551}

\bibitem[{{Rickett}(1970)}]{ric70}
{Rickett}, B.~J. 1970, \mnras, 150, 67, \dodoi{10.1093/mnras/150.1.67}

\bibitem[{{Sand} {et~al.}(2023){Sand}, {Breitman}, {Michilli}, {Kaspi},
  {Chawla}, {Fonseca}, {Mckinven}, {Nimmo}, {Pleunis}, {Shin}, {Andersen},
  {Bhardwaj}, {Boyle}, {Brar}, {Cassanelli}, {Cook}, {Curtin}, {Dong}, {Eadie},
  {Gaensler}, {Kaczmarek}, {Lanman}, {Leung}, {Masui}, {Rahman}, {Pandhi},
  {Pearlman}, {Petroff}, {Rafiei-Ravandi}, {Scholz}, {Shah}, {Smith}, {Stairs},
  \& {Stenning}}]{sbm+23}
{Sand}, K.~R., {Breitman}, D., {Michilli}, D., {et~al.} 2023, arXiv e-prints,
  arXiv:2307.05839, \dodoi{10.48550/arXiv.2307.05839}

\bibitem[{{Taylor}(1992)}]{tay92}
{Taylor}, J.~H. 1992, Philosophical Transactions of the Royal Society of London
  Series A, 341, 117, \dodoi{10.1098/rsta.1992.0088}

\bibitem[{{Virtanen} {et~al.}(2020){Virtanen}, {Gommers}, {Oliphant},
  {Haberland}, {Reddy}, {Cournapeau}, {Burovski}, {Peterson}, {Weckesser},
  {Bright}, {van der Walt}, {Brett}, {Wilson}, {Millman}, {Mayorov}, {Nelson},
  {Jones}, {Kern}, {Larson}, {Carey}, {Polat}, {Feng}, {Moore}, {VanderPlas},
  {Laxalde}, {Perktold}, {Cimrman}, {Henriksen}, {Quintero}, {Harris},
  {Archibald}, {Ribeiro}, {Pedregosa}, {van Mulbregt}, \& {SciPy 1.0
  Contributors}}]{vgo+20}
{Virtanen}, P., {Gommers}, R., {Oliphant}, T.~E., {et~al.} 2020, Nature
  Methods, 17, 261, \dodoi{10.1038/s41592-019-0686-2}

\end{thebibliography}

\end{document}